\documentclass[fleqn,usenatbib]{mnras}
\usepackage{newtxtext,newtxmath}
\usepackage[T1]{fontenc}
\DeclareRobustCommand{\VAN}[3]{#2}
\let\VANthebibliography\thebibliography
\def\thebibliography{\DeclareRobustCommand{\VAN}[3]{##3}\VANthebibliography}

\usepackage{graphicx}

\usepackage{amsmath}
\usepackage{amssymb}
\newcommand{\vpeak}{V_{\rm peak}}
\newcommand{\vmax}{V_{\rm max}}
\newcommand{\mpeak}{M_{\rm peak}}
\newcommand{\minfall}{M_{\rm infall}}
\newcommand{\fk}{f_{\rm k}}

\newcommand{\hMsun}{ h^{-1}{\rm M_{ \odot}}}
\newcommand{\hMpc}{ h^{-1}{\rm Mpc}}

\newcommand{\ihMpcC}{ h^{3}{\rm Mpc}^{-3}}

\newcommand{\sig}{\sigma_{8}}
\newcommand{\OmM}{\Omega_\mathrm{M}}
\newcommand{\Omb}{\Omega_{\rm b}}
\newcommand{\h}{h}
\newcommand{\ns}{{n_{\rm s}}}
\newcommand{\Mnu}{M_{\rm \nu}}
\newcommand{\wa}{w_{\rm a}}
\newcommand{\wz}{w_{0}}
\newcommand{\OmMh}{\Omega_\mathrm{M}h^2}

\newcommand{\sigL}{\sigma_{\rm lum}}
\newcommand{\tmerger}{t_{\rm merger}}
\newcommand{\Fk}{f_{\rm k,cen=sat}}
\newcommand{\betaL}{\beta_{\rm lum}}
 
\newcommand{\proj}{${w_{\rm p}}$}
\newcommand{\mono}{$\xi_{\ell=0}$}
\newcommand{\quadr}{$\xi_{\ell=2}$}
\newcommand{\hexa}{$\xi_{\ell=4}$}
 
\newcommand{\Mr}{$M_{\rm r}$}
\newcommand{\MrMax}{M_{\rm r}^{\rm max}}
\newcommand{\Mra}{$n_{_{M_r\ -19.5}}$}
  
\newcommand{\Mrc}{$n_{_{M_r\ -20.5}}$}
\newcommand{\Mrd}{$n_{_{M_r\ -21.0}}$}
\newcommand{\Mre}{$n_{_{M_r\ -21.5}}$}

\newcommand{\rmin}{r_{\rm min}}

\title[Inferring cosmology and assembly bias from galaxy clustering]{The MillenniumTNG Project: Inferring cosmology from galaxy clustering with accelerated N-body scaling and subhalo abundance matching}

\author[S. Contreras et al.]{%
Sergio Contreras$^{1}$\thanks{E-mail: sergio.contreras@dipc.org},
Raul E. Angulo$^{1,2}$\thanks{E-mail: reangulo@dipc.org},
Volker Springel$^{3}$,
Simon D. M. White$^{3}$,
Boryana Hadzhiyska$^{4,5,6}$,
\newauthor%
Lars Hernquist$^{4}$,
R\"udiger Pakmor$^{3}$,
Rahul Kannan$^{4}$,
C\'esar Hern\'andez-Aguayo$^{3,7}$,
Monica Barrera$^{3}$,
\newauthor%
Fulvio Ferlito$^{3}$, 
Ana Maria Delgado$^{4}$,
Sownak Bose$^{8}$, and Carlos Frenk$^{8}$
\\%
\\%
$^{1}$Donostia International Physics Center (DIPC), Donostia-San Sebastian, Spain\\%
$^{2}$IKERBASQUE, Basque Foundation for Science, 48013, Bilbao, Spain\\%
$^{3}$Max-Planck-Institut f\"{u}r Astrophysik, Karl-Schwarzschild-Str. 1, 85748, Garching, Germany\\%
$^{4}$Harvard-Smithsonian Center for Astrophysics, 60 Garden St, Cambridge, MA 02138, USA\\%
$^{5}$Miller Institute for Basic Research in Science, University of California, Berkeley, CA, 94720, USA\\%
$^{6}$Physics Division, Lawrence Berkeley National Laboratory, Berkeley, CA 94720\\%
$^{7}$Excellence Cluster ORIGINS, Boltzmannstrasse 2, 85748 Garching, Germany\\%
$^{8}$Institute for Computational Cosmology, Department of Physics, Durham University, South Road, Durham, DH1 3LE, UK
}

\date{Accepted XXX. Received YYY; in original form ZZZ} 

\pubyear{2023}

\begin{document}
\label{firstpage}
\pagerange{\pageref{firstpage}--\pageref{lastpage}} 
\maketitle

% Abstract of the paper
\begin{abstract}
We introduce a novel technique for constraining cosmological parameters and galaxy assembly bias using non-linear redshift-space clustering of galaxies. We scale cosmological N-body simulations and insert galaxies with the SubHalo Abundance Matching extended (SHAMe) empirical model to generate over 175,000 clustering measurements spanning all relevant cosmological and SHAMe parameter values. We then build an emulator capable of reproducing the projected galaxy correlation function at the monopole, quadrupole and hexadecapole level for separations between $0.1\,\hMpc$ and $25\,\hMpc$. We test this approach by using the emulator and Monte Carlo Markov Chain (MCMC) inference to jointly estimate cosmology and assembly bias parameters both for the MTNG740 hydrodynamic simulation and for a semi-analytical galaxy formation model (SAM) built on the MTNG740-DM dark matter-only simulation, obtaining unbiased results for all cosmological parameters. For instance, for MTNG740 and a galaxy number density of $n\sim 0.01 \ihMpcC$, we obtain $\sig=0.799^{+0.039}_{-0.044}$ and $\OmMh= 0.138^{+ 0.025}_{- 0.018}$ (which are within 0.4 and 0.2 $\sigma$ of the MTNG cosmology). For fixed Hubble parameter ($h$), the constraint becomes $\OmMh= 0.137^{+ 0.011}_{- 0.012}$. Our method performs similarly well for the SAM and for other tested sample densities. We almost always recover the true amount of galaxy assembly bias within one sigma. The best constraints are obtained when scales smaller than $2\,\hMpc$ are included, as well as when at least the projected correlation function and the monopole are incorporated. These methods offer a powerful way to constrain cosmological parameters using galaxy surveys.
\end{abstract}

%%%%%%%%%%%%%%%%%%%%%%%%%%%%%%%%%%%%%%%%%%%%%%%%%%

%%%%%%%%%%%%%%%%% BODY OF PAPER %%%%%%%%%%%%%%%%%%

\section{Introduction}
\label{sec:intro}

The distribution of galaxies in the Universe reflects two different factors: (i) the cosmological context, which determines how dark matter haloes cluster, and (ii) galaxy formation physics, which governs how different galaxies populate dark matter haloes. It is not easy to disentangle these two effects when analysing the clustering of galaxies selected by directly observable properties.  Despite this, forward modelling can be used to constrain cosmological information using galaxy clustering.

By modelling galaxy clustering for a particular cosmology and comparing the results to observational data, one can ascertain a model's realism.  This comparison can in turn be used to constrain cosmology using observed galaxy clustering. The resulting constraints will vary according to the precision of the galaxy clustering model, and the amount of cosmological information contained in the observational clustering measurements used.

A straightforward method for constraining cosmology with galaxies is to examine their clustering at comparatively large scales. On these scales, modelling galaxies is much simpler than at small scales, as there is no need to know the distribution of galaxies within haloes in detail. Additionally, the available observational galaxy samples that reach these large scales contain only a small fraction of satellite galaxies, typically having only one galaxy per halo.

While procedurally convenient, ignoring the clustering information from smaller scales (e.g.~\citealt{DonaldMcCann:2021}) leads to a weakening of the recoverable constraints. To include these scales, however, one needs a model that can reproduce the distribution of galaxies inside haloes. The Halo Occupation Distribution model (HOD,  \citealt{Jing:1998a,Benson:2000,Peacock:2000,Berlind:2003,Zheng:2005,Zheng:2007, C13,Guo:2015a,C17}) is one of the most widely used approaches for addressing this issue. The HOD quantifies the average number of galaxies ($\langle N \rangle$) that populate a halo as a function of its mass ($M_{h}$). The HOD can be used to predict galaxy clustering in two main ways: (a) by combining it with an analytic halo clustering model such as the ``halo model'' (e.g. \citealt{Guzik:2001}) or another similar approach; or (b) by populating the haloes of an N-body dark matter simulation. 

HOD models with an analytic clustering prescription for the non-linear power spectrum are often used in the literature to constrain cosmological information. \cite{Tinker:2012} combined the projected correlation function (\proj) with the mass-to-galaxy-number ratio show that galaxy clustering can be used to constrain cosmological parameters such as $\OmM$ and $\sig$. Similarly, \cite{Cacciato:2013} (see also \citealt{vandenBosch:2013,More:2013}) constrained these same cosmological parameters using the halo model and the conditional luminosity function. 

Mock catalogues based on HOD modelling applied to dark matter simulations have also  been used to measure cosmological parameters. Thus, \cite{Reid:2014} inferred $f\sig$ from the CMASS sample of BOSS (see also \citealt{Lange:2019,Lange:2021}) and by using a set of 40 simulations from the {\sc AEMULUS} project, \cite{Zhai:2019} showed the constraining power of a 7-parameter HOD for determining $f\sig$, $\OmM$ and $\sig$. More recently, \cite{Yuan:2022} used the {\sc AbacusSummit} suite of simulations to constrain $f\sig$, $\OmM$ and $\sig$. While successful, the HOD approach has some important limitations. Modern implementations require a large number of free parameters (up to 12 in e.g.,~\citealt{Yuan:2021a}) to produce realistic galaxy clustering measurements. Such a large number of free parameters, and their possible degeneracies with cosmological parameters, may limit  the cosmological constraints these models can achieve.

Another method for reproducing galaxy clustering is to populate the subhaloes of an N-body simulation using a subhalo abundance matching technique (SHAM, e.g. \citealt{Vale:2006, Conroy:2006}). This method is based on the assumption that the most massive subhaloes contain the most massive and luminous galaxies. The SHAM technique requires a better resolved dark matter simulation than needed for the HOD approach. Additionally, it necessitates the computation of subhalo properties that involve the subhaloes' merger trees, such as their peak subhalo mass ($\mpeak$), or their maximum circular velocity ($\vpeak$). The primary advantage of this technique is that it can reproduce galaxy clustering realistically both at small and large scales, even when only one free parameter is used \citep{ChavesMontero:2016}. 

\cite{Simha:2013} used the SHAM technique to constrain the values of $\OmM$ and $\sig$ from the SDSS. They scaled a single high-resolution dark matter simulation using the procedure described in \cite{Angulo:2010} in order to generate corresponding realizations of different cosmologies. This method modifies the results of a given simulation in order to replicate the properties and mass distribution of another model with a different cosmology, without having to actually simulate this different model. By fitting the projected correlation function, the authors were able to constrain the SDSS cosmological parameters to an uncertainty of less than 10\%.

In this paper, we extend the work of \cite{Simha:2013} by using updated versions of the scaling and SHAM techniques to constrain cosmological parameters of  SDSS-like samples with errors $\sim 5\%$.  In \cite{C20} (see also \citealt{Zennaro:2019, Angulo:2021, Arico:2021, Ondaro:2021}) we have improved the precision of the scaling technique by including an additional correction for the matter distribution on large scales and for the 1-halo term. We have demonstrated that we can scale dark matter simulations to within 3\% accuracy for the the matter, halo, and subhalo power spectra. For populating the subhaloes, we use the SubHalo Abundance Matching extended model (SHAMe, \citealt{C21c}). The SHAMe model generalizes the basic SHAM by including orphans, tidal disruption, and a flexible amount of galaxy assembly bias. These additions enhance the predictions of galaxy clustering in both real- and redshift-space. Also, in this form the model can be run with simulations of intermediate resolution, whereas the standard SHAM requires higher resolution simulations to reproduce the clustering of high number density samples. 

These improved techniques enable us to scale dark matter simulations across a wide range of cosmologies and to populate them with SHAMe mocks. We then use these mocks to build an emulator capable of predicting galaxy clustering statistics both rapidly and precisely (a few milliseconds per query) as a function of SHAMe and cosmological parameters. This in turn allows us to use standard MCMC approaches to constrain the cosmological paramaters of a given galaxy clustering dataset, either from observation or from an  independent simulation.

In this work, we make use of the MillenniumTNG (MTNG) simulation project to test and validate our inference pipeline. The recently introduced MTNG project \citep[see][]{HernandezAguayo2022, Pakmor2022, Barrera2022, Hadzhiyska2022a, Hadzhiyska2022b, Kannan2022, Bose2022, Ferlito2022, Delgado2022} combines the largest volume, high-resolution hydrodynamic simulation of galaxy formation to date (MTNG740, a $500\,h^{-1}{\rm Mpc} \simeq 740\,{\rm Mpc}$ periodic box) with a sequence of matching dark matter-only simulations, as well as with simulations that include massive neutrinos. Importantly, the MTNG simulations have been run independently, in fact with different codes and with different cosmological settings, from the simulations we use to build our emulator. Testing with an independent hydrodynamic simulation of known cosmology yields a powerful challenge for  the accuracy and robustness of our approach.  In addition, to further test the validity of our method, we also us it to infer cosmological paramters from a galaxy catalogue  besed on a semi-analytical model (SAM) of galaxy formation applied to one of the MTNG dark matter only simulations \citep{Barrera2022}.

When applying the MCMC approach to these simulated catalogues, we use their predictions for the galaxy clustering in real- and redshift-space. The extent to which obtained unbiased and accurate estimates of the true cosmological parameters of MTNG provides a strong test of the power of our methodology. Furthermore, the flexible level of assembly bias allowed in the SHAMe model, allows us to test if we can measure the true level of assembly bias in  MTNG when marginalising over the cosmological parameters. To the best of our knowledge, this makes our study the first work to constrain galaxy assembly bias using a SHAM-like approach without assuming the cosmology of the target sample (several groups have recently achieved this using the HOD framework, e.g., \citealt{Zhai:2022,Lange:2022,Yuan:2022}).

The outline of this paper is as follows: Section~\ref{sec:simulations} presents the dark matter simulations and the galaxy population models. The computation of the galaxy clustering and the quantification of errors in our model are presented in Section~\ref{sec:clustering}. The main results of this paper, the constraints on the MTNG cosmological parameters from galaxy clustering in the MTNG740 hydrodynamic simulation and in the MTNG740-DM+SAM, are shown in  Section~\ref{sec:constrain}. We then turn to discussing our constraints on the level of galaxy assembly bias in Section~\ref{sec:AB}. Finally, we give a summary and discussion of our findings in Section~\ref{sec:Summary}.

Unless otherwise stated, the standard units in this paper are $\hMsun$ for masses, $\hMpc$ for distances, and ${\rm km\,s^{-1}}$ for the velocities. Magnitudes are in all cases absolute magnitudes, and refer to the rest frame. All logarithm values are in base 10.

\section{Numerical simulations and galaxy population models}
\label{sec:simulations}

In this section, we first describe the suite of dark matter-only simulations we employ to create our mocks in Section~\ref{sec:bacco}. In Section~\ref{sec:scaling}, we briefly introduce the scaling technique applied to  our N-body simulations. In Section~\ref{sec:shame}, we present the galaxy clustering model (SHAMe) we use to populate the (scaled) simulations with galaxies. Finally, we describe the MTNG hydrodynamic simulation (Section~\ref{sec:tng}) as well as the MTNG semi-analytical model (Section~\ref{sec:sam}) that yield our target galaxy samples for testing and validating our inference technique.

\subsection{The Bacco simulations}
\label{sec:bacco}

\begin{table}
    \centering
        \caption{The number densities used in this work, along with the equivalent cuts in  \Mr~for  each galaxy formation model and the SHAMe model. Notice that the cuts for the SHAMe model are similar to the ones of \citet{Guo:2015} for the SDSS. This is because the SHAMe used a luminosity function from the SDSS \citep{Blanton:2001}. To facilitate further comparisons, we named each number density based on the cut value in the SDSS.}
    \begin{tabular}{ccccccccc}
        \hline
        Name & $ n$ &  $\MrMax$ & $\MrMax$ & $\MrMax$ & $\MrMax$\\
         & ${\rm 10^{-3}}\ihMpcC$ &  SDSS & SHAMe & MTNG & SAM\\
        \hline
        \Mra & 11.64  & -19.5 & -19.39 & -19.81 & -20.77 \\
        \Mre & 6.37   & -20.0 & -19.98 & -20.86 & -21.53 \\
        \Mrc & 3.13   & -20.5 & -20.48 & -21.76 & -22.09 \\
        \Mrd & 1.16   & -21.0 & -20.97 & -22.60 & -22.64 \\
        \Mre & 0.29   & -21.5 & -21.49 & -23.44 & -23.16 \\
        \hline
    \end{tabular}
    \label{table:nden}
\end{table}

\begin{table}
    \centering
        \caption{Cosmological parameters of the five main simulation sets used in this work. The vilya, nenya, narya and power cosmologies are used in the construction of our inference methodology, whereas the MTNG cosmology is used exclusively to test the performance of our method. All these simulations have values of $\Mnu=0$, $\wz=-1$, and $\wa=0$. }
    \begin{tabular}{cccccccccc}
        \hline
        Cosmology &  $\sig$ &  $\OmM$ & $\Omb$ & $\h$ & $\ns$\\
        \hline
        MTNG  & 0.8159 & 0.3089 & 0.0486 & 0.6774 & 0.9667\\
        vilya  & 0.9 & 0.270 & 0.060 & 0.65 & 0.92\\
        nenya  & 0.9 & 0.315 & 0.050 & 0.60 & 1.01\\
        narya  & 0.9 & 0.360 & 0.050 & 0.70 & 1.01\\
        power  & 0.9 & 0.3071 & 0.0483 & 0.6777 & 0.9611\\
        \hline
    \end{tabular}
    \label{table:sims}
\end{table}

Four pairs of dark matter-only simulation were used to construct the emulator of this study: "vilya", "nenya", "narya", and "power". The cosmological parameters of these simulations (see Table~\ref{table:sims}) were chosen to minimize the error of the scaling technique  (following \citealt{C20}, see Section~\ref{sec:scaling} for more details). These paired simulations were run with opposite initial Fourier phases, using the procedure of \cite{Angulo:2016} that suppresses cosmic variance by up to 50 times compared to a random simulation of the same volume. Each simulation has a volume of $(512 \hMpc)^3$, similar to the $(500 \hMpc)^3$ of the MTNG, and a resolution of $1536^3$ particles.

These simulations, as well as all simulations specifically run for this work,  were carried out with an updated version of {\tt L-Gadget3} \citep{Angulo:2012}, a `lean' (particularly memory-efficient) version of {\tt GADGET} \citep{Springel:2005}. This code was also used to run the Millennium-XXL simulation and the Bacco Simulations \citep{Angulo:2021}. Using a Friend-of-Friend algorithm \citep[{\tt FOF}][]{Davis:1985} with a linking length of 0.2, and an extended version of {\tt SUBFIND} \citep{Springel:2001}, this version of the code allows an on-the-fly identification of haloes and subhaloes. In particular, our updated version of {\tt SUBFIND} can better identify substructures by considering the information of its past history, while also measuring properties that are non-local in time, such as the peak halo mass ($\mpeak$), peak maximum circular velocity ($\vpeak$), infall subhalo mass ($\minfall$), and mass accretion rate, among others. 

Additionally, we make use of two suites of simulations to evaluate the scaling technique's performance. The first set of simulations consists of 33 paired simulations of $1536^3$ particles and a box length of $\sim 512\, \hMpc$. Except for one cosmological parameter, which we change to a different value for each pair of simulations, these simulations have the ``nenya cosmology''. These simulations are used to measure the dependence of the galaxy clustering on cosmology (see Section~\ref{sec:param_space} for more details).

The second additional suite of simulations  consists of two sets of 15 simulations that were run with volumes of  $(256\ \hMpc)^3$ and $768^3$ particles (same resolution as the previous simulations, but lower volume), and were used to quantify the error of the scaling technique (see Section~\ref{sec:SclErr} for more details). The cosmological parameters of these simulations are chosen from a Latin-Hypercube over the range of parameters we cover with the scaling technique (see Section~\ref{sec:param_space} for the details of the hyper-parameter cover). We changed $\sig$, $\OmM$, $\Omb$, $\h$, and $\ns$  in the first set of 15 simulations, whereas we varied $\Mnu$, $\wa$, and $\wz$ in addition to the five previous parameters for the second set.

\subsection{The scaling technique}
\label{sec:scaling}
The scaling technique \citep{Angulo:2010} modifies the outputs of a dark matter simulation by displacing its particles, haloes and subhaloes. Its goal is to produce a matter distribution that is comparable to that produced by a simulation run with a different cosmology. Numerous studies have established the method's accuracy (e.g. \citealt{Ruiz:2011,Guo:2013a,Zennaro:2019}). More recently, \cite{C20} achieved a 3\% precision in reproducing the matter, halo, and subhalo power spectra over a wide range of cosmological parameter space. This was accomplished by scaling simulations centred on three distinct cosmologies. The cosmological parameter space covered by \citet{C20} is similar to the one of this work (eqs. 2-9, see Section~\ref{sec:param_space} for more details), which is approximately 10$\sigma$ around Planck's best-fitting values (see section~3 of \citealt{C20} for a detailed explanation on how these  parameters were chosen). The three main cosmologies used in \cite{C20} are the same as the cosmologies of our ``vilya'', ``nenya'' and ``narya'' simulations. However, we found that by including an additional simulation (``power''),  we can reduce the error on the power spectrum to   $\sim 2\%$. As a point of reference, reading and scaling one of our simulations takes approximately 15 seconds when using a single CPU.

The scaling technique enabled us to concentrate our computational resources on running highly resolved N-body simulations on the four cosmologies mentioned previously, rather than on tens of low-resolution simulations for a variety of cosmologies. High-resolution simulations are needed to run more realistic mock algorithms, like the one we describe in the next section.

\subsection{The subhalo abundance matching extended model}
\label{sec:shame}
To constrain cosmological information from galaxy clustering, we require a model capable of realistically and efficiently populating dark matter simulations. To accomplish this, we employ the {\bf S}ub-{\bf H}alo {\bf A}bundance {\bf M}atching {\bf e}xtended model (SHAMe) developed by \cite{C21c}. The two primary advantages of this model are: (a) the small number of free parameters, and (b) the precision with which it reproduces galaxy clustering in real- and redshift-space, particularly on small scales. The small number of free parameters reduces the susceptibility to degeneracy with cosmological parameters. The high accuracy on small scales is key to making proper use of the constraining power of galaxy clustering in the nonlinear regime.

Just as in the standard SHAM approach \citep[][]{Vale:2006, Conroy:2006,Reddick:2013,C15,ChavesMontero:2016,Lehmann:2017,Dragomir:2018,Hadzhiyska:2021b}, our model begins by matching a subhalo property (in this case, $\vpeak$) to the expected luminosity function. We define $\vpeak$ as the maximum circular velocity ($\vmax \equiv {\rm max}\sqrt{GM(<r)/r}$) achieved over the evolution of a halo/subhalo. We use the luminosity function of \citet{Blanton:2001}, which facilitates further comparison with observational data. As mentioned in \cite{C21c}, when galaxies are selected using number density cuts, the choice of a specific luminosity function has little to no effect on the galaxy clustering statistics.

After constructing the basic SHAM, the model introduces orphan galaxies; i.e.~satellite structures with known progenitors that the simulation cannot resolve but are expected to exist in the halo. We do this by following the most bound particle of the subhalo after we can no longer identify it. We assume that an orphan merges with its central structure when the time since accretion exceeds a dynamical friction timescale, $t_{\rm infall} > t_{\rm d.f.}$, where $t_{\rm d.f.}$ is the dynamical friction time computed at the moment the satellite subhalo become an orphan
and using a modified version of Eq. 7.26 of \cite{BT:1987},
\begin{equation}
t_{\rm d.f.} = \dfrac{1.17\ \tmerger\ d_{\rm host}^2\ V_{\rm host} (M_{\rm host}/10^{13}\ h^{-1}\mathrm{M}_{\odot})^{1/2}}{G \ln(M_{\rm host}/M_{\rm sub}+1)\ M_{\rm sub}},
\end{equation}
\noindent where $\tmerger$ is a free dimensionless parameter that effectively regulates the number of orphan galaxies; $d_{\rm host}$ is the distance of the subhalo to the centre of its host halo; $V_{\rm host}$ is the virial velocity of the host halo; $M_{\rm host}$ is the virial mass of the host halo, and $M_{\rm sub}$ is the subhalo mass. 

Next, galaxies that became satellites a long time ago are removed from the sample. After a period of time, satellite galaxies begin to lose stellar mass, reducing their luminosity. Additionally, satellite galaxies lose their cold gas, which reddens the galaxies and reduces their luminosity in certain bands (including \Mr). By excluding all galaxies that have been satellites for an extended period of time, i.e. $t_{\rm infall} > \betaL\, t_{\rm dyn}$, with $t_{\rm dyn}$ the halo's dynamical time, defined as $0.1/H(z)$ and $\betaL$ being a free parameter, we can improve the galaxy clustering predictions. We also tested alternative approaches, such as removing substructures using their lost subhalo mass (as in \citealt{C21a} and \citealt{Moster:2018}) and other more complex approaches, but we found that our simple approach fits the galaxy clustering the best for a luminosity-selected galaxy sample.

The final step in the SHAMe implementation is to include additional galaxy assembly bias. Galaxy assembly bias \citep{Croton:2007} is the change in galaxy clustering caused by the propagation of halo assembly bias \citep{Gao:2005, Gao:2007} into the galaxies. This propagation occurs because the occupation of galaxies depends on halo properties that cause halo assembly bias (e.g.~occupancy variations, \citealt{Zehavi:2018, Artale:2018}). To this date, to our knowledge, there has been no absolute confirmation of the (non)existence of this kind of assembly bias for real galaxies.
The level of assembly bias in a hydrodynamic simulation is not necessarily the same as in a SHAM \citep{ChavesMontero:2016} or in a SAM \citep{C21c,Hadzhiyska:2021}. Additionally, none of these coincides necessarily with the level of assembly bias in the real Universe. To account for the uncertainty surrounding the assembly bias of the target galaxy sample, we introduce a tuneable level of such bias in our model galaxy samples. While cosmology has a negligible effect on assembly bias, excluding it could potentially introduce a systematic bias in our constraints on cosmological parameters \citep{C21b}.

To introduce variable assembly bias into our samples, we follow the procedure developed by \cite{C21a} which utilises the individual bias-per-object of the galaxies \citep{Paranjape:2018} to choose preferentially more/less biased objects. In a nutshell, the model exchanges the luminosities of galaxies with similar values of $\vpeak$, to make their luminosities correlate/anticorrelate with lare-scale environment density (see also \citealt{Hadzhiyska:2020,Xu:2021a,Xu:2021b} for other studies that look at the impact of environment on other galaxy population models). We preserve the satellite fraction of the original galaxy sample by performing this step independently for central and satellite galaxies. Thus, the method uses two free parameters to control the level of galaxy assembly bias, $f_{\rm k,cen}$ and $f_{\rm k,sat}$, for central and satellite galaxies, respectively. A value of $f_{\rm k}=1$ (-1) means a maximum (minimum) galaxy assembly bias signal, while a value of 0 means the same assembly bias level as a standard SHAM. For simplicity, during this work, we set $f_{\rm k,cen} = f_{\rm k,sat}$. We check that this approximation has a low impact on the cosmological and assembly bias constraints. We did this by looking at the posteriors of the cosmological parameters on models with one and two assembly bias parameters, and found no significant difference in the constraints. Note that this assembly bias implementation allows us to constrain the level of assembly bias from the galaxy clustering of the galaxy formation models, without assuming any cosmology for the target sample. 

One effect that is not completely covered by the SHAMe model but is well known to affect galaxies is velocity bias (e.g. \citealt{Guo:2015, Ye:2017}). It has been reported that ignoring the velocity bias effect of central galaxies can potentially bias cosmological constraints \citep{Lange:2022, Yuan:2022, Zhai:2022}. The SHAMe model partially accounts for this effect by following the positions and velocities of the subhalos rather than the haloes. When fitting the galaxy clustering, we also take the SHAMe model error into account (see \S~\ref{sec:SHAMeErr} for more details). Nonetheless, in the future, we will explore incorporating velocity bias into the SHAMe to improve the model's accuracy even further.

\subsection{The MTNG740 simulation}
\label{sec:tng}

To validate our inference methodology, we make use of galaxy samples from the MillenniumTNG simulations \citep[see][for the introductory papers to the project]{HernandezAguayo2022, Pakmor2022, Barrera2022, Kannan2022, Bose2022, Hadzhiyska2022a, Ferlito2022, Delgado2022}. These calculations are meant to extend the two well-known Millennium \citep{Springel:2005} and IllustrisTNG \citep{TNGa, TNGb, TNGc, TNGd, TNGe, Nelson:2019a, Nelson:2019b, Pillepich:2019} projects in a direction that allows accurate studies of the galaxy-halo connection and the impact of baryonic physics on clustering, in particular, to much larger cosmological volumes than previously possible. To this extent, the project includes a new state-of-the-art hydrodynamical simulation with the galaxy formation model of IllustrisTNG \citep{Pillepich:2018,Weinberger:2017}, but carried out in a $(500\,h^{-1}{\rm Mpc})^3$ volume, the size of the original Millennium simulation, hence the name MillenniumTNG for the project (or MTNG for short).

A full overview of the simulation set of MTNG can be found in \citet{HernandezAguayo2022}. Besides large hydrodynamical simulations, it also includes a series of dark matter-only simulations, and runs that explicitly include massive neutrinos. We will here focus on the flagship hydrodynamical model, MTNG740, which is based on the IllustrisTNG physics implementation but offers a volume nearly 15 times larger than TNG300, the biggest box of IllustrisTNG, at slightly poorer mass resolution.  The simulated volume is a periodic box with $500\hMpc\simeq 740\,{\rm Mpc}$ on a side; the number of dark matter particles and gas cells is  each $4320^3$, implying an average gas cell mass of $2.00\times10^7\,\hMsun$ and a mass resolution of $1.12\times 10^8\,\hMsun$ for the dark matter. The simulation adopted cosmological parameters identical to IllustrisTNG for ease of comparison, consistent with \citet{Planck2015}\footnote{$\OmM$ = 0.3089, $\Omb$ = 0.0486, $\sig$ = 0.8159, $\ns$ = 0.9667 and $h$ = 0.6774. }. The initial conditions for MTNG were made by fixing initial power mode amplitudes to their expected {\it rms} \citep{Angulo:2016}, which significantly reduces the effect of cosmic variance on a simulation, at least for second-order statistics. Only one simulation has been run, however, not a full pair, due to the very large computational cost. For the corresponding dark matter-only simulations (see below), a pair with reflected phases has been simulated, however, in order to enable the full reduction in large-scale statistical ``noise''.

MTNG740 was run using the moving-mesh code \texttt{AREPO} \citep{AREPO} and accounts for radiative cooling and star formation in the gas, the growth of supermassive black holes, as well as associated energetic feedback processes from supernovae and black holes, among other processes of galaxy formation physics. Galaxies are produced as agglomerations of star particle with properties that can be directly measured from the simulation. (See \citet{Pakmor2022} for an analysis of how well basic properties of the galaxy population of MTNG740 agree with observations and with the previous simulations of IllustrisTNG.) The dark-matter only version of MTNG740 (MTNG740-DM) was run with the {\tt Gadget-4} code \citep{Gadget-4}.

To build our target galaxy sample, we select the most luminous galaxies in the $r$-band at $z=0.1$. We define a galaxy's luminosity as the sum of the luminosities of all its stellar particles. We chose the most luminous galaxies down to number densities of $11.64$, $6.37$, $3.13$, $1.16$, and $0.29\times 10^{-3}\ihMpcC$, respectively. These number densities are the same as the ones chosen by \cite{Guo:2015}, who computed the SDSS observational galaxy clustering. Using these number densities will facilitate a direct comparison with cosmological and galaxy assembly bias constraints we plan to obtain for the SDSS galaxy clustering in future work. 

We show the equivalent luminosity cuts for each of our number densities in Table~\ref{table:nden}. While the magnitude cuts differ from those used in  SDSS, galaxy clustering is more sensitive to the number density itself than to the value of individual property cuts, making comparisons between samples with the same number density more appropriate for these types of studies \citep{C13}. While MTNG740 is similarly successful as IllustrisTNG in reproducing a large number of observables \citep{Pakmor2022}, note that we do not require our target simulation to be completely realistic. We are here more concerned with testing our methodology's ability to recover the cosmology and galaxy formation information of any underlying galaxy model that is fed to it, regardless of whether the model agrees with observation or not.

\subsection{The semi-analytical model}
\label{sec:sam}

To further validate our procedures and to test their constraining capacity, we also use them to analyse a galaxy sample derived from a semi-analytical model for galaxy formation (SAM, e.g. \citealt{Kauffmann:1993, Cole:1994, Somerville:1999,  Somerville:2008, Bower:2006, Lagos:2008, Benson:2010, Benson:2012, Jiang:2014, Croton:2016, Lagos:2018, Stevens:2018, Henriques:2020}), applied to MTNG740-DM-2-A \citep[see Table 1 of][for the specifications of the simulations]{HernandezAguayo2022}, one of the dark matter-only companion runs of MTNG740. Unlike the hydrodynamic simulations, the baryonic matter is here not simulated alongside the dark matter, but rather is tracked using simplified analytical modelling grafted on top of stored subhalo merger trees created from a dark-matter-only simulation. This approach enables easy and rapid examination of alternative galaxy formation assumptions, because it allows them to be varied without requiring a new and computationally expensive dynamical simulation.

We here use the new semi-analytic methodology developed by \citet{Barrera2022} for application to MTNG, as realized in the most recent version of the {\small L-GALAXIES} code. This model  is based on a long history of prior development of the ``Munich semi-analytic model'' \citep{White:1991, Kauffmann:1993, Kauffmann:1999, Springel:2001,Springel:2005, DeLucia:2004, Croton:2006, DeLucia:2007, Guo:2011, Guo:2013, Henriques:2013,Henriques:2020}. The new version of \citet{Barrera2022} has substantially improved tracking of subhalos and galaxies over cosmic time, which, in particular, allows continuous outputing on the past light cone. In terms of physics modeling, however, it largely relies on the parameterization of \cite{Henriques:2015}.

Apart from using the SAM catalogue of \citet{Barrera2022}, we also produced four additional models ourselves with extreme galaxy formation parameter variations, in order to test the robustness of our inference pipeline to significant modifications of galaxy formation physics. These models were implemented for a dark matter-only simulation that used the same cosmology, initial conditions, and volume as the MTNG740-DM simulation, but has a lower resolution ($1536^3$ particles). Also, these models were run with the older public version of {\small L-GALAXIES}\footnote{\url{https://lgalaxiespublicrelease.github.io}}, using (mostly) its default parameter set. To produce extreme model variations, we changed the supernova energy efficiency parameter by multiplying/dividing it by a factor of 10 or 100. We opted to change this parameter because it was the one that maximally influenced the clustering of \Mr~selected galaxy samples at fixed number density.

\section{Galaxy clustering as a function of cosmology}
\label{sec:clustering}

We aim to constrain cosmology through galaxy clustering by combining the scaling technique for generating dark matter simulations in various cosmologies and the SHAMe technique for populating these simulations with galaxies. We will use $r$-band selected galaxies from the MTNG hydro simulation, and a SAM based on a MTNG dark matter-only simulation as target samples to test our inference capability of cosmological parameters. As previously mentioned, we employ the $r$-band to facilitate future applications to observational data.

In this section, we build a Monte Carlo approach that systematically compares the clustering predictions of our mocks with the clustering of our target samples, providing joint constraints on both SHAMe and cosmological parameters. In Section~\ref{sec:param_space}, we detail the cosmological parameter space we explore. In Section~\ref{sec:emulator}, we build an emulator that predicts galaxy clustering as a function of our SHAMe and cosmological parameters. Finally, Section~\ref{sec:Err} details the covariance matrix used to run our Monte-Carlo approach. 

\subsection{The parameter space}
\label{sec:param_space}

\begin{figure*}
\includegraphics[width=0.95\textwidth]{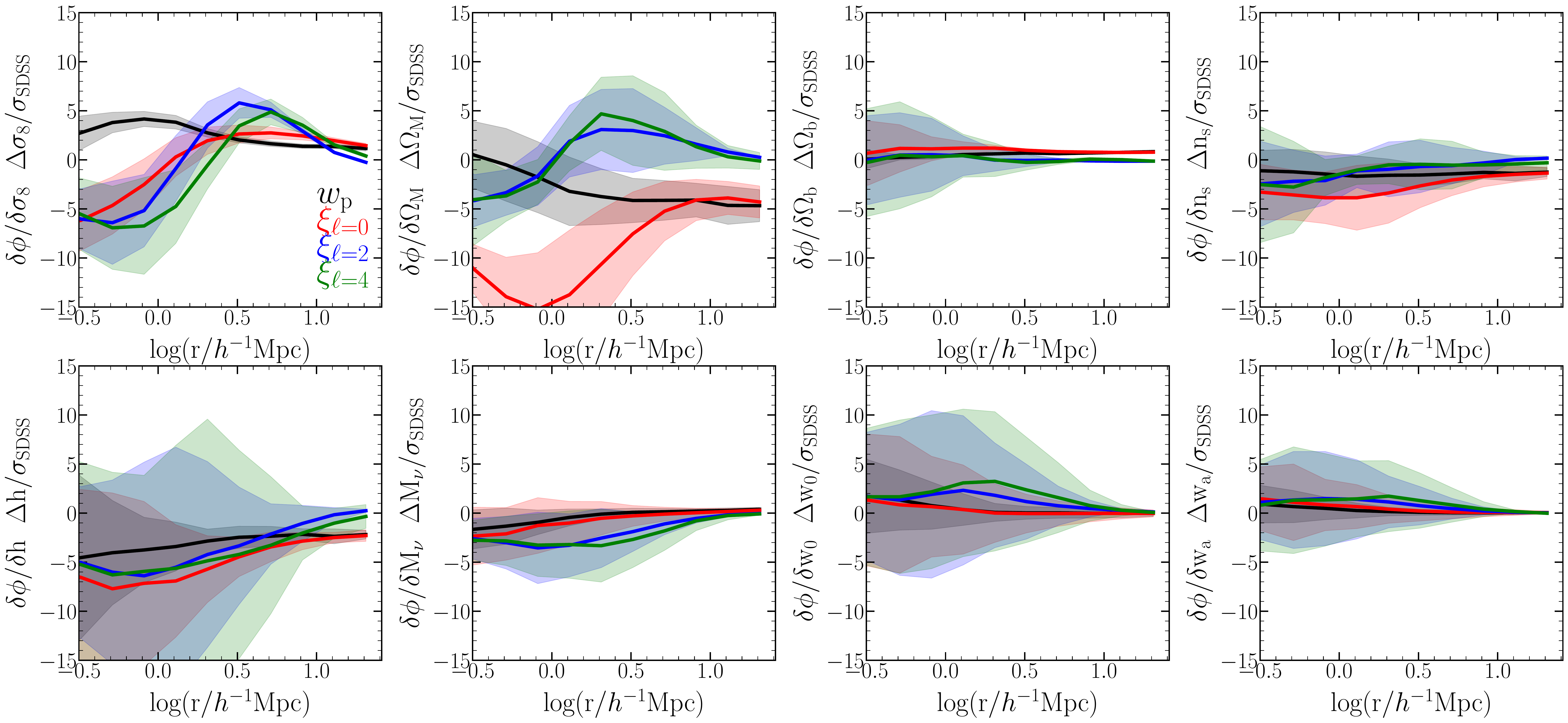}
\caption{The relative change in galaxy clustering (\proj~in black, \mono~in red, \quadr~in blue, and  \hexa~in green) as a function of scale for different cosmological parameters: $\sig$, $\OmM$, $\Omb$, $\h$, $\ns$, $\Mnu$, $\wa$, and $\wz$. The change in clustering is computed as the clustering variation for galaxy samples run in simulations with one different cosmological parameter($\delta \phi$), normalized by the variation on that cosmological parameter ($\delta {\rm param}$). This change in clustering is then scaled by the maximum change in the cosmological parameter (which is normally 10 $\sigma$ around the Planck best fit cosmology, $\Delta {\rm param}$) and normalized again by the expected error of the SDSS on each clustering statistics ($\sigma_{\rm SDSS}$). The solid lines represent the mean of 4 different measurements, while the shaded regions cover one standard deviation around the mean value. (See Section~\ref{sec:param_space} for more details.)}
\label{Fig:derivative}
\end{figure*}

The range of cosmological parameters we can explore is limited by the performance of the scaling technique.  \cite{C20} demonstrate that we can scale to a parameter space greater than $10\sigma$ around the Planck best fit cosmology based on simulations with just three parameter sets, those of our ``vilya'', ``nenya'' and ``narya'' simulations. As mentioned previously, we added an additional simulation, ``power'', to further improve the accuracy of the scaling technique. The range of cosmologies we looked at are:

\begin{eqnarray}
\label{eq:par_range}
\sig                      &\in& [0.65, 0.9]\\
\OmM                      &\in& [0.23, 0.4]\\
\Omb                      &\in& [0.04, 0.06]\\
\ns                       &\in& [0.92, 1.01]\\
h\,[100\,{\rm km}\,{\rm s^{-1}} {\rm Mpc^{-1}}]  &\in& [0.6, 0.8]\\
\Mnu\,[{\rm eV}]        &\in& [0.0, 0.4]\\
\wz                     &\in& [-1.15, -0.85]\\
\wa                     &\in& [-0.3, 0.3]
\end{eqnarray}
Notice that the lower limit of $\sig$ is 0.65, lower than the limit set by \cite{C20} of 0.73. We recently found that increasing the range over which we scale this property has no discernible effect on the scaling technique's error.

As one might expect, not all cosmological parameters have the same effect on galaxy clustering. To quantify this, we employ the suite of 33 paired simulations described in Section~\ref{sec:bacco}. Each simulation has $1536^3$ particles and a box size of $\sim512\ \hMpc$. The simulations have a similar cosmology to the ``nenya'' simulation, except that one parameter is varied within the specified range given above. We populate all these simulations using the SHAMe model, using the same parameters in every case for the \Mra density sample. The SHAMe parameters are those that minimise the clustering difference between MTNG740 and a SHAMe mock run over a MTNG dark matter-only simulation. 
%We have also populated the simulations with a basic SHAM technique and found similar results.

We quantified the variation in the projected correlation function (\proj), monopole (\mono), quadrupole (\quadr), and hexadecapole (\hexa) of the correlation function between 5 consecutive simulations by varying only one single cosmological parameter. For example, for $\OmM$ we compute the differences between the  simulations with $\OmM = \{0.23, 0.27\}$,  $\OmM = \{0.27, 0.315\}$,  $\OmM = \{0.315, 0.36\}$, and  $\OmM = \{0.36, 0.4\}$. The corresponding change in the statistics ($\delta \phi$) is divided by the change in each of the parameters (in this case, $\delta \OmM$) to get a derivative of the clustering statistics with respect to the cosmological parameter. To account for the significance of this derivative, we then normalize by the ratio between the total change of the parameter (e.g.~for $\OmM$ this would be $0.4-0.23 = 0.17$) and the error expected from SDSS for that statistic ($\sigma_{\rm SDSS}$, see~Section~\ref{sec:SDSSErr}). These normalized derivatives for the clustering statistics are shown in Fig.~\ref{Fig:derivative}.

The galaxy clustering proved to be more sensitive to changes in $\OmM$, $\sig$ and $\h$ than to any of the other parameters in our set. For convenience, we thus restrict ourselves in the following to constraining these three parameters. We also tested including the neutrino mass $\Mnu$, which also displays some clustering dependence, but not unexpectedly, we were not able to constrain it or find any meaningful relation worth reporting here.

We would like to emphasize that other clustering statistics, such as the 3PCF \citep{Guo:2016b} or the kNN-CDF \citep{Banerjee:2020}, may be more sensitive to changes in other cosmological parameters and should not be ruled out as properties that can be constrained at these scales via galaxy clustering. A more in-depth examination of these dependencies will be conducted in  future work.

\subsection{Emulating the galaxy clustering}
\label{sec:emulator}

As described thus far, we have developed a method capable of efficiently creating mock galaxy catalogues at any redshift. As a reference, for a single CPU, the time required to: (1) read a pair of dark matter simulations; (2) scale the simulations to a target cosmology; (3) create 4 distinct mocks, each with a unique random seed, and (4) compute the \proj,~\mono,~\quadr,~and~\hexa~for three different lines of sight, for 6 different number densities (this last part done with 4 CPUs), take less than 7 minutes. While reasonably fast, this is still too slow for a MCMC-like approach.

To speed up the generation of clustering predictions, we thus developed an emulator based on $\sim 175,000$ clustering measurements of mocks with varying cosmologies and SHAMe parameters. The emulator was constructed using a feed-forward  neural network in a manner similar to that described in  \cite{Angulo:2021} and \cite{Arico:2021}. The architecture used consists of two fully connected hidden layers with 200 neurons each, and a rectified linear unit activation function for the projected correlation function and monopole of the correlation function, as well as three layers for the quadrupole and hexadecapole, with each statistic being represented by an independent network. We have also tested other configurations reaching similar performances. 

The neural networks were trained using the Keras front-end of the Tensor-flow library \citep{tensorflow}. We used the Adam optimization algorithm, with a learning rate of 0.001, and a mean squared error loss function. We split our dataset into disjoint groups for training and validation. The training set contains 90\% of the data and required approximately 45 minutes of processing per number density/statistic on a single Nvidia Quadro RTX 8000 GPU card. Evaluating the four emulators takes $\sim 47$ milliseconds on a laptop, with $\sim 0.5$ seconds to evaluate 100,000 samples (it is more efficient to evaluate the data in larger groups). As part of this paper, we are also making this emulator publicly available\footnote{\url{http://www.mtng-project.org}}.

\subsection{Error quantification } 
\label{sec:Err}
In this section, we look at the different uncertainties associated with our model. To account for them, we create a covariance matrix that includes all of these possible error sources. By omitting any systematic error, we risk biasing our predictions. In Section~\ref{sec:SDSSErr}, we examine the contribution of cosmic variance to the error. We quantify the errors introduced by the SHAMe model, the scaling technique, and the emulator in Sections~\ref{sec:SHAMeErr},~\ref{sec:SclErr}, and \ref{sec:EmuErr}, respectively. Finally, in Section~\ref{sec:errsum}, we show how these errors can be combined, and how much they contribute to the total error.

\subsubsection{SDSS errors}
\label{sec:SDSSErr}
Since the initial conditions of the MTNG were produced by fixing the initial power spectrum, and the initial conditions of the Bacco simulations were also run with the fixed \&  paired method \citep{Angulo:2016}, we do not expect a significant contribution of cosmic variance for the scales we are interested in ($r<25\hMpc$). Nonetheless, in order to replicate cosmological constraints realistically, we assume an error in our clustering prediction comparable to that of the SDSS. We use the covariance matrix provided by \cite{Guo:2015} (see also \citealt{Zehavi:2011}). As previously stated, the number densities of our samples are identical to those of SDSS, and their mean redshift ($z=0.1$) is comparable as well.  This particular selection of galaxy samples thus facilitates any future study of the SDSS clustering.

\cite{Guo:2015} computed the covariance matrix ($\rm C_{\rm v,sdss}$) using 400 jackknife samples \citep{Zehavi:2002,Norberg:2008} measuring the projected correlation function (\proj) and the multipoles of the correlation function (\mono,~\quadr,~and~\hexa). The resulting covariance matrix also contains the cross term between these statistics. All of their clustering measurements were made under the assumption of a Planck cosmology.

Because we computed the clustering of the target sample (MTNG-hydro and MTNG-SAM) at $z=0.1$, and the distance units include an $h$ factor, they should be mostly unaffected by the Alcock-Paczynski effect \citep{AlcockPaczynski}. We demonstrate this by performing some extreme changes in cosmology and discovering that they are mostly negligible, consistent with the findings of \cite{Guo:2015}.

\begin{figure*}
\includegraphics[width=0.95\textwidth]{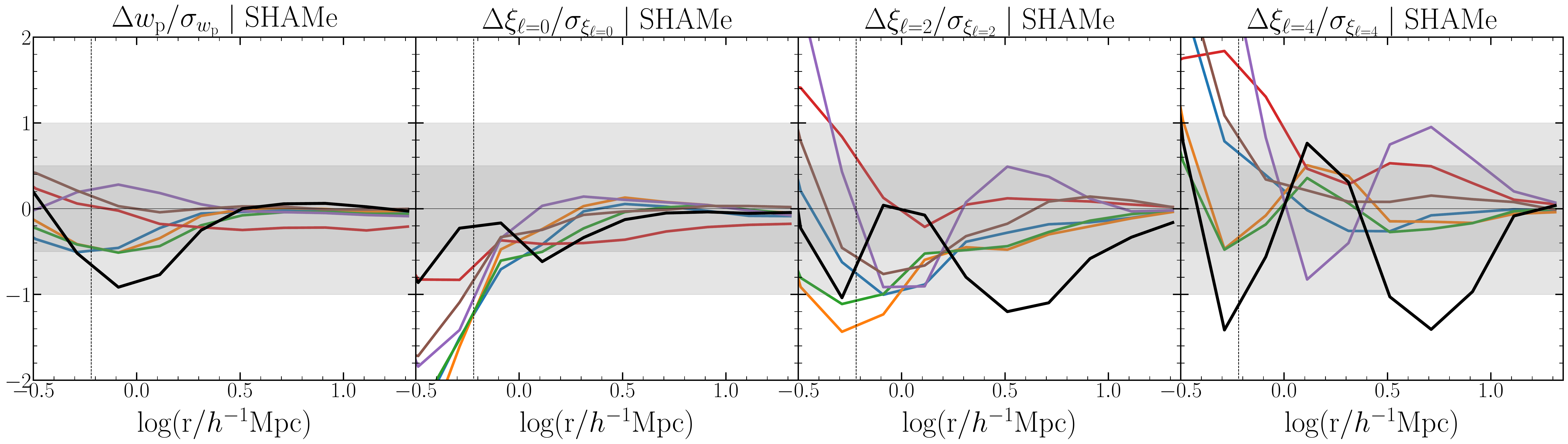}
\caption{The errors of the SHAMe model for the \Mra\ galaxy sample. The black line indicates the difference between the MTNG's galaxy clustering and the best fitting SHAMe. The coloured lines depict the same difference, but for various SAMs rather than the MTNG. The vertical line shows the minimum scale used in the fitting. The difference is normalized by the SDSS error, also for the \Mra galaxy sample. }
\label{Fig:SHAMe_err}
\end{figure*}

\subsubsection{Galaxy modelling errors}
\label{sec:SHAMeErr}

To quantify the error associated with the SHAMe model, we compare the clustering of the MTNG740 simulation and the five SAMs run with different (and extreme) physical parameters to the clustering of a SHAMe mock run over a MTNG dark matter-only simulation. We use the Particle Swarm Optimization algorithm {\tt PSOBACCO}\footnote{\url{https://github.com/hantke/pso_bacco}}, described in \cite{PSOBACCO}, to determine the SHAMe model parameters that best fit the MTNG hydro run and the SAMs. The fits are performed by minimizing the $\chi^2$ computed using the covariance matrix of the SDSS for scales greater than a given $r_{\rm  min}$. We generate several covariance matrices for each of the $r_{\rm  min}$ values used in this paper.

We compute a covariance matrix from the differences in clustering between mocks and galaxy formation models:
\begin{equation}
\begin{split}
{\rm C_{\rm v,SHAMe }}({\boldsymbol V}_{i},{\boldsymbol V}_{j}) =  
\dfrac{1}{N} \sum_{l=1}^{N} ({\boldsymbol V}_{\rm SHAMe}-{\boldsymbol V}_{\rm gal.\ form.})_i^l \\({\boldsymbol V}_{\rm SHAMe}-{\boldsymbol V}_{\rm gal.\ form.})_{j}^l,
\end{split}
\end{equation}
\noindent with ${\rm C_{\rm v,SHAMe }}$ being the covariance matrix from the SHAMe modelling, ${\boldsymbol V}_{\rm SHAMe}$ and ${\boldsymbol V}_{\rm gal.\ form.}$ representing the clustering vector (\proj,~\mono,~\quadr,~and~\hexa) of the mocks and the galaxy formation models, respectively. Fig.~\ref{Fig:SHAMe_err} shows the ratio between the clustering measurements of the galaxy formation models and the SHAMe for an $r_{\rm min} = 0.6\, \hMpc$. The differences considered here are between the thick solid line and the black horizontal line (i.e. the difference in the clustering of the galaxy formation model) and not between the mean of the distribution of the individual lines. These differences agree with the ones found for stellar mass-selected galaxies \citep{C21c}.

\begin{figure*}
\includegraphics[width=0.95\textwidth]{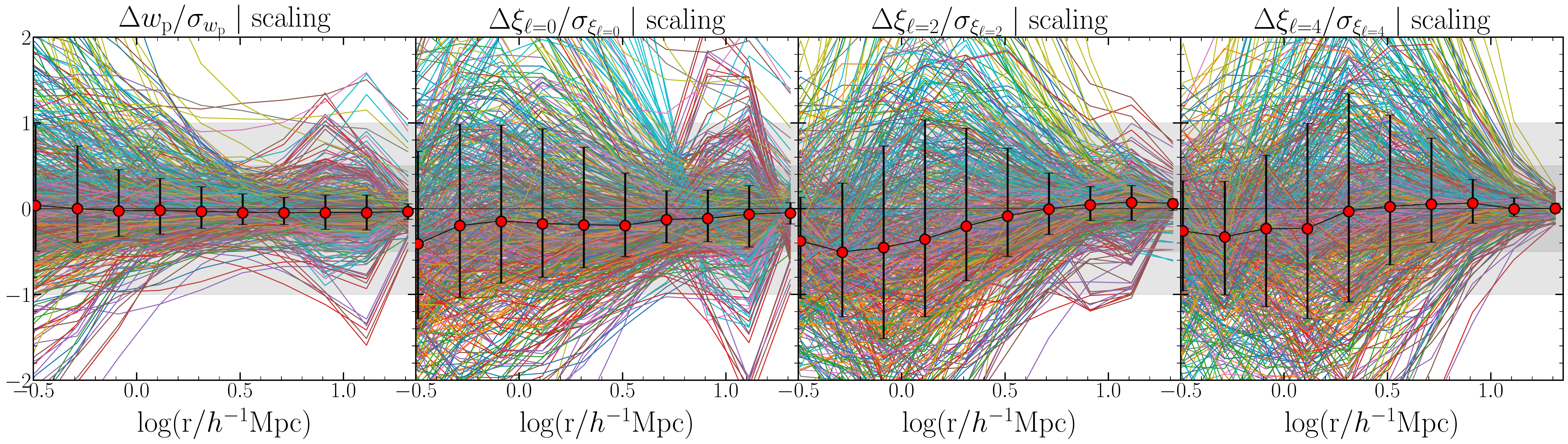}
\caption{The error of the scaling technique. The lines depict the differences in galaxy clustering between 600 SHAMe models run over 30 dark matter simulations, and those same 600 models run over scaled simulations. The differences are normalised by the SDSS error estimated for a comparable galaxy sample. The red circles and error bars represent the distribution's median, and $16^{\rm th}$ and $84^{\rm th}$ percentiles, respectively.}
\label{Fig:Scale_err}
\end{figure*}

\subsubsection{Scaling errors}
\label{sec:SclErr}

We now look at the error introduced by using scaled N-body simulations. For this, we use the suite of 30 paired simulations with different cosmologies described in Section~\ref{sec:bacco}. We divide this suite into two groups. In the first group we varied $\sig$, $\OmM$, $\Omb$, $\ns$ and $h$, and in the second group we varied the neutrino mass ($\Mnu$) and the dark energy equation-of-state parameters $\wz$ and $\wa$ in addition  to the parameters varied in the first group. For each simulation, we compute the galaxy clustering for 20 SHAMe mocks, each with different and randomly selected parameters.

The projected correlation function, monopole, quadrupole, and hexadecapole of the mocks run on full N-body simulations are compared to those run on scaled simulations for the \Mra sample in Fig.~\ref{Fig:Scale_err}. At all scales, the scaling error is subdominant for the projected correlation function. For the monopole of the correlation function, the error due to the scaling technique only becomes relevant for small scales. The errors of the quadrupole and hexadecapole are more significant, comparable to the ones of the SDSS at scales below $2\,\hMpc$. On larger scales, the scaling can successfully predict all statistics. Other number densities exhibit similar trends.

Similarly to the previous section, we construct a covariance matrix from the differences between the scaled and real galaxy clustering measurements,
\begin{equation}
{\rm C_{\rm v,scl }}({\boldsymbol V}_{i},{\boldsymbol V}_{j}) =\dfrac{1}{N} \sum_{l=1}^{N} ({\boldsymbol V}_{\rm scl}-{\boldsymbol V}_{\rm target})_i^l ({\boldsymbol V}_{\rm scl}-{\boldsymbol V}_{\rm target})_{j}^l,
\end{equation}
\noindent with ${\rm C_{\rm v,scale }}$ being the covariance matrix from the scaling, ${\boldsymbol V}_{\rm target}$ being the clustering vector (\proj,~\mono,~\quadr~and~\hexa) from the two groups of simulations, and ${\boldsymbol V}_{\rm scl}$ representing the clustering vector from the scaled simulations. 

\begin{figure*} 
\includegraphics[width=0.95\textwidth]{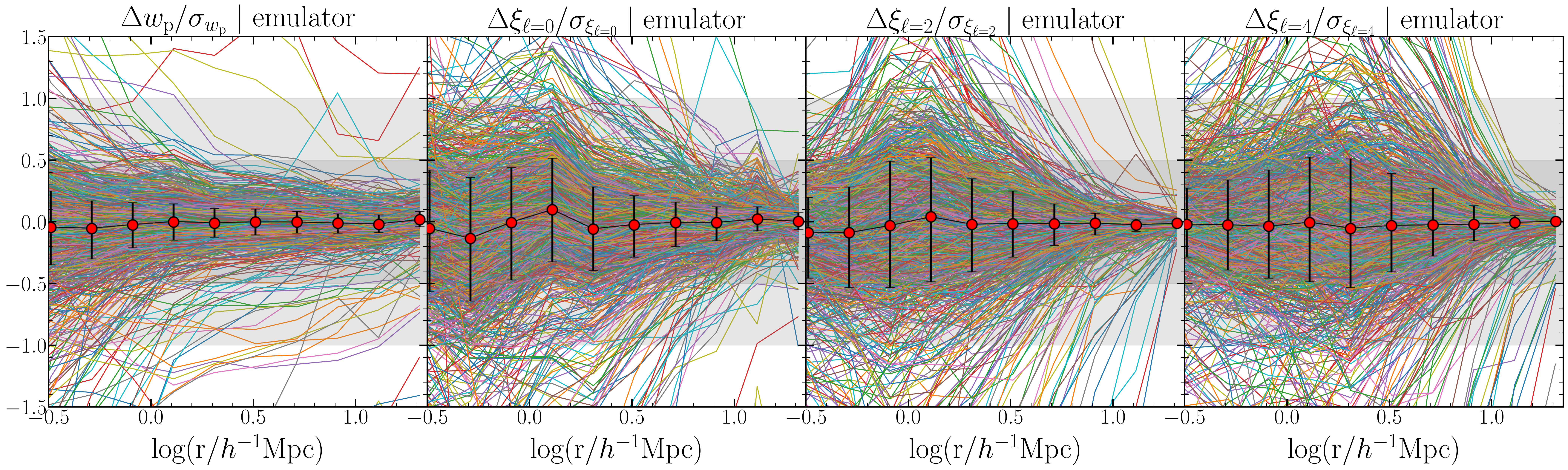}
\caption{The error of the emulator. The lines depict the differences in galaxy clustering between 1,000 scaled SHAMe mocks and their emulator counterparts. The differences are normalised by the SDSS error estimated for a comparable galaxy sample. The red circles and error bars represent the distribution's median, and $16^{\rm th}$ and $84^{\rm th}$ percentiles.}
\label{Fig:Emu_Err} 
\end{figure*}

\subsubsection{Emulator errors}
\label{sec:EmuErr}

To assess the emulator's accuracy, we compare its clustering predictions against a subsample of 1,000 scaled simulations that were not used in the emulator's training or testing. We then construct the covariance matrix, ${\rm C_{\rm v, emulator}}$, computed as:
\begin{equation}
{\rm C_{\rm v,emu }}({\boldsymbol V}_{i},{\boldsymbol V}_{j}) =\dfrac{N-1}{N} \sum_{l=1}^{N} ({\boldsymbol V}_{\rm emu}-{\boldsymbol V}_{\rm val})_i^l ({\boldsymbol V}_{\rm emu}-{\boldsymbol V}_{\rm val})_{j}^l,
\end{equation}  
\noindent with ${\boldsymbol V}_{\rm emu}$ and ${\boldsymbol V}_{\rm val}$ representing the combined clustering data (\proj,\ \mono,\ \quadr, and \hexa) for the emulated and validation data, respectively. 

In Fig.~\ref{Fig:Emu_Err}, we compare the clustering of mocks and the emulator for the \Mra density sample. The emulator's overall performance is good, with the lowest source of errors shown thus far. Lower number densities have a slightly greater dispersion, but they are always the lowest source of errors.

\begin{figure*}
\includegraphics[width=0.95\textwidth]{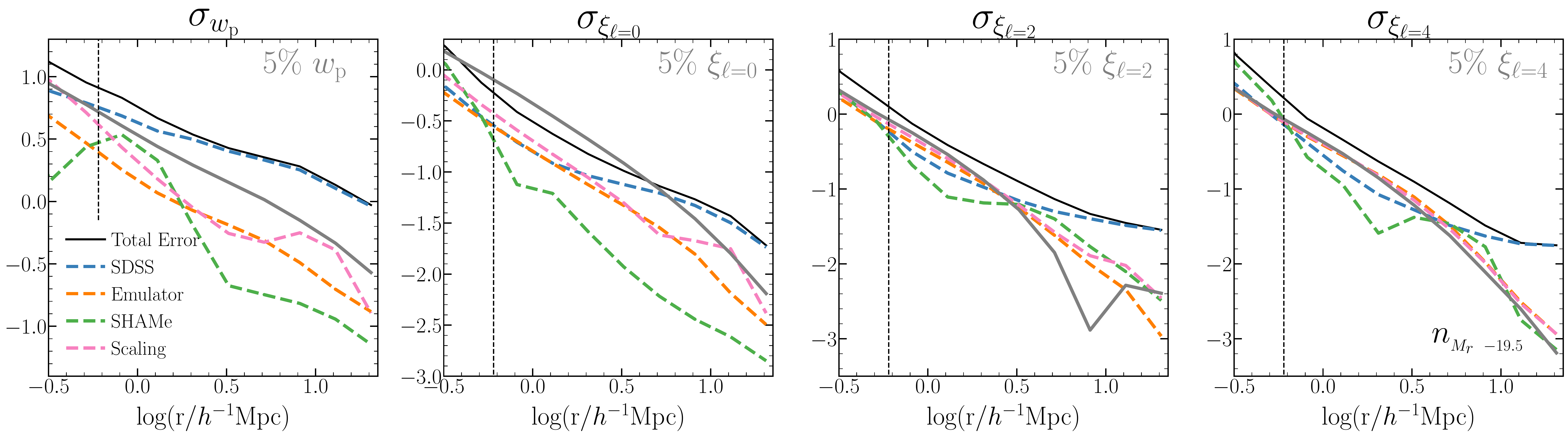}
\caption{The contribution of all errors to galaxy clustering as a function of scale for the \Mra\  sample. The vertical line indicates the minimum scale at which the SHAMe mocks were fitted, $\rmin=0.6\,\hMpc$. We highlight in greys  5\% of the value of each clustering statistic as a point of reference. If the errors are all independent, the total error (black line)  equals the sum of all sources of errors (coloured dashed lines). The primary sources of errors are the assumed cosmic variance errors from the SDSS, and the errors of the scaling technique. }
\label{Fig:full_err}
\end{figure*} 

\begin{figure*}
\includegraphics[width=0.95\textwidth]{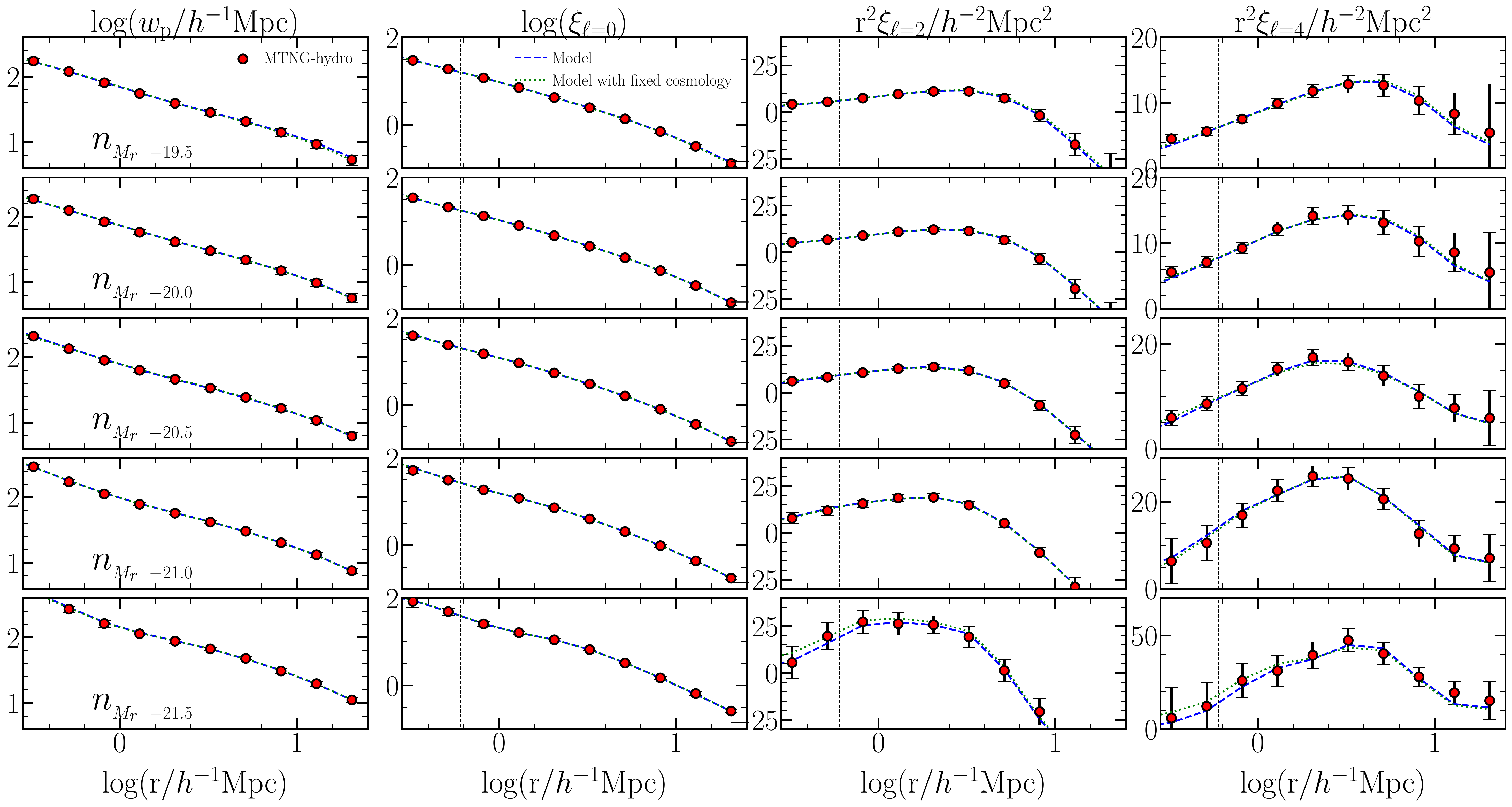}
\includegraphics[width=0.95\textwidth]{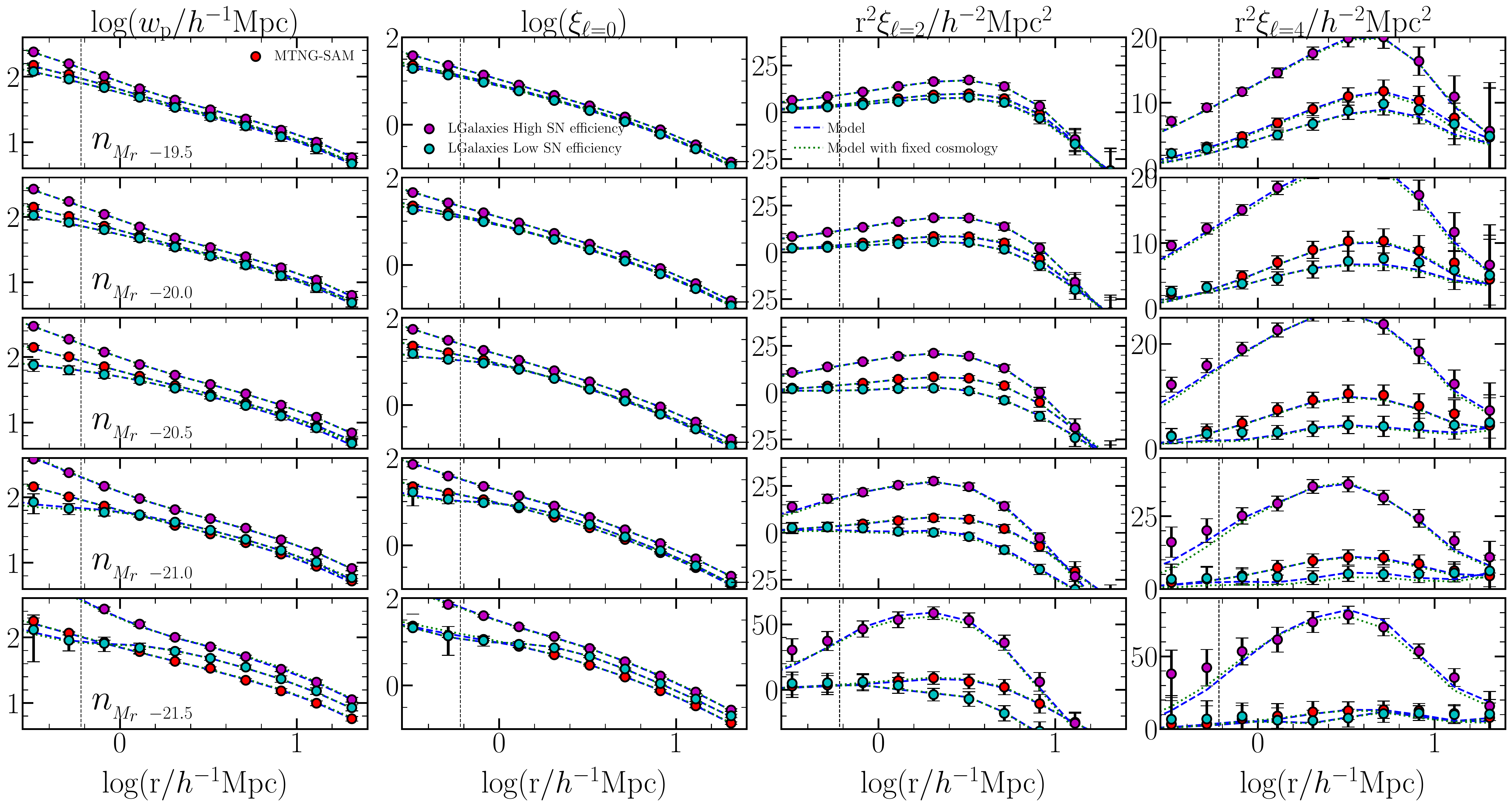}
\caption{{\it Top panel:} The galaxy clustering for the MTNG (red circles); the emulated clustering of the best fitting SHAMe mock fixing the cosmology to the one of the MTNG (green dotted line) and allowing any cosmology (blue dashed line). The error bars in the MTNG clustering represent the model's total error (see Section~\ref{sec:errsum} for more details). The vertical line indicates the fitting's minimum scale. {\it Bottom panel:} Similar to the top panel, but for three SAMs with different physical prescriptions.}
\label{Fig:full_fit}
\end{figure*}

\subsubsection{Combining all errors}
\label{sec:errsum}

After examining these possible sources of error, we now combine them to obtain the model's total error. Given that the errors are mostly uncorrelated, it is reasonable to assume that the total error can be simply described as the sum of all individual errors:
\begin{equation}
{\rm C_{\rm v,tot }} = {\rm C_{\rm v,SDSS }} + {\rm C_{\rm v,SHAMe }} + {\rm C_{\rm v,scl }} + {\rm C_{\rm v,emu }}.
\end{equation}

We show in Fig.~\ref{Fig:full_err} the square root of the diagonal of the covariance matrices of all the sources of error discussed so far, for the \Mra density sample. As expected, the SDSS covariance matrix is the primary source of error. The scaling technique is the second most significant source of error, which dominates at scales below $\sim 2\, \hMpc$ for the monopole, quadrupole, and hexadecapole. In Section~\ref{sec:con_cosmo}, we will show how these two major sources of errors affect our cosmological constraints.

Fig.~\ref{Fig:full_fit} shows the emulated clustering of the best fitting SHAMe mock when the cosmology is fixed to the one of MTNG740 (green dotted line) and when any cosmology is allowed (blue dashed line). The top panel compares the emulator's clustering predictions to the MTNG740 galaxies, while the bottom panel comparesto three different semi-analytic runs. 
%\textcolor{red}{What are the hi- and lo-SN runs here? Earlier you mentioned two models of each type.}
The fits are made by minimizing the $\chi^2$ computed using the combined covariance matrix at scales above $0.6\, \hMpc$.  Utilizing the entire covariance matrix may result in the best fit not passing through the centre of each data point. This could account for some of the quadrupole and hexadecapole deviations. Nonetheless, the fits perform admirably with a $\chi^2/{\rm d.o.f.} < 1$. In the following section, we will use this emulator to assess the technique's constraining power when SHAMe and cosmological parameters are varied simultaneously.

\section{Constraining cosmology using galaxy clustering}
\label{sec:constrain}

In this section, we use our emulator to constrain cosmological and galaxy formation information from the  galaxy clustering of the MTNG740 simulation and the fiducial SAM catalogue based on MTNG740-DM. Using a Monte-Carlo Markov Chain (MCMC), we obtain posterior distributions for our parameters. We use the public code {\tt emcee} \citep{emcee} employing 1000 chains with an individual length of 10,000 values, and a burn-in phase of 1,000. While perhaps atypical, this combination of chains and steps is ideal for an emulator-based MCMC, which is extremely efficient when computing multiple points simultaneously. We test additional combinations of MCMC parameters and obtain nearly identical results in all cases. The average computing time of each MCMC analysis was $\sim 20$ minutes. We computed the likelihood function as:
\begin{equation}
ln\ \mathcal{L} = -\chi^2/2,
\end{equation}
\noindent with $\mathcal{L}$ being the likelihood and $\chi^2$ computed as
\begin{equation}
\chi^2 = ({\boldsymbol V}_{\rm emu\ mock} - {\boldsymbol V}_{\rm gal.\ form.})^{\rm T}{\rm C_{\rm v,tot }}^{-1}({\boldsymbol V}_{\rm emu\ mock} - {\boldsymbol V}_{\rm gal.\ form.}),
\end{equation}
where ${\rm C_{\rm v, tot }}$  is the covariance matrix, and ${\boldsymbol V}_{\rm emu\ mock}$ and ${\boldsymbol V}_{\rm gal.\ form.}$ represent the clustering vector (\proj, \mono, \quadr,~and \hexa) of the emulator and the galaxy formation models, respectively.  The maximum number of free parameters is 8 (4 SHAMe parameters and 4 cosmological parameters).  Due to the emulator's efficiency, we can easily test different covariance matrix configurations and different minimum scales for the galaxy clustering, and quantify the effect of the different parameters on the resulting constraints.

\begin{figure*}
\includegraphics[width=0.95\textwidth]{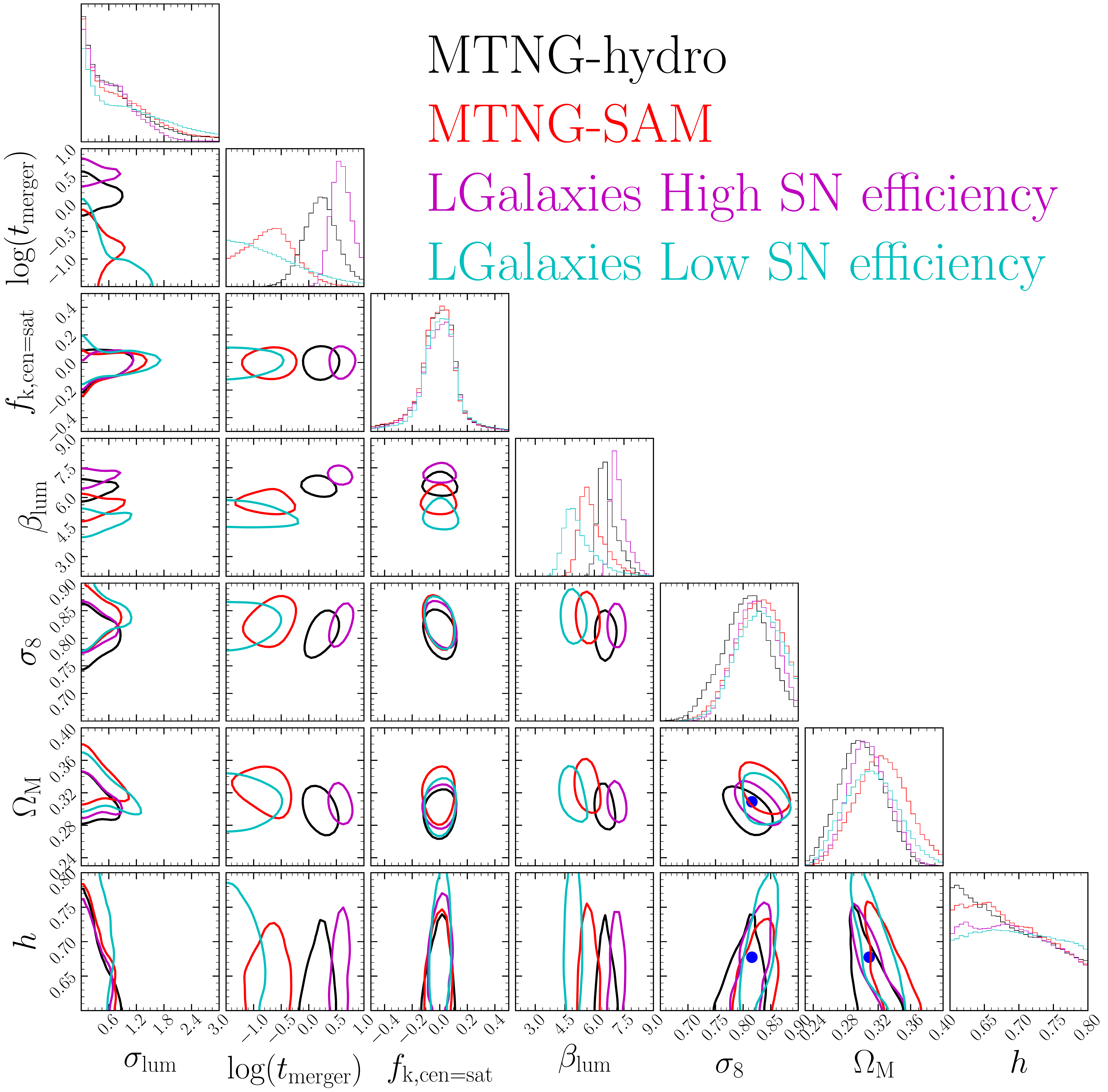}
\caption{Marginalized 1$\sigma$ confidence regions for 4 SHAMe  parameters and three cosmological parameters derived from the galaxy clustering (\proj, \mono, \quadr, and \hexa) of the MTNG740 simulation (black line) and different SAMs (red, cyan \& magenta lines) for a sample with a number density of \Mra $= 11.64\ 10^{-3}\ \ihMpcC$. The probability distribution function for each parameter is displayed at the top of each column. The blue circles represent the simulations' true cosmology.}
\label{Fig:full_mcmc}
\end{figure*}

\begin{figure}
\includegraphics[width=0.45\textwidth]{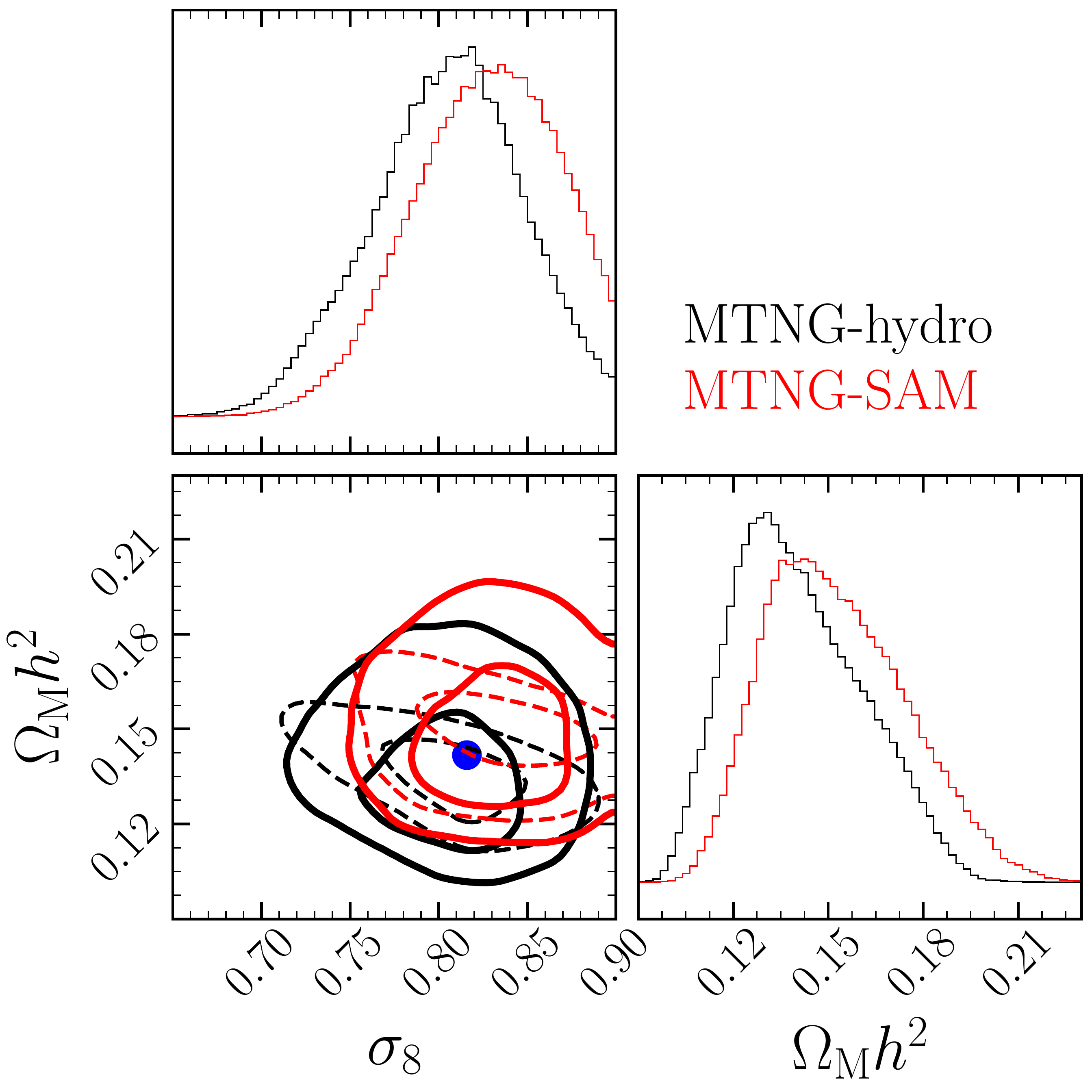}
\caption{Marginalized $1\,\sigma$ and  $2\,\sigma$ credibility regions for $\sig$ and $\OmMh$ for the \Mra\ density sample. The dashed line contours represent the marginalized region when fixing the Hubble parameter to its correct value ($h=0.6774$). For visual clarity, the top histograms are only shown for the non-fixed $\h$ runs.}
\label{Fig:simple_mcmc}
\end{figure}

In Fig.~\ref{Fig:full_mcmc}, we show the $1\,\sigma$ confidence regions for the cosmological and SHAMe parameters when fitting \proj,~\mono,~\quadr,~and~\hexa~for a \Mra density sample. The blue dot represents the galaxy formation models' cosmology. While the MTNG-SAM and MTNG-hydro model exhibit some differences in their confidence regions, the correct cosmology is recovered within one sigma in both cases. We notice that we are unable to capture the entire distribution of the Hubble parameter ($\h$), which means that this parameter cannot be constrained using these clustering statistics.  This is in part because there is a degeneracy between $\h^2$ and $\OmM$. To account for this, we show in Fig.~\ref{Fig:simple_mcmc} the $1\,\sigma$ and   $2\,\sigma$ confidence regions for $\sig$ and $\OmMh$, marginalized over the other parameters, for the same galaxy samples as above. The parameters look to be constrained within the parameter space explored, with the correct cosmology recovered within one sigma. When the Hubble parameter is set to the correct value, the constraints for $\OmMh$ become significantly tighter (as expected), but we observe no significant improvement in the constraints on $\sig$. For this number density, we can place the following constraints, $\sig\ = 0.799^{+0.039}_{-0.044}$ and $0.826^{+0.041}_{-0.045}$, and $\OmMh= 0.138^{+ 0.025}_{- 0.018}$ and  $0.151^{+ 0.026}_{- 0.020}$ for the MTNG-hydro and the MTNG-SAM galaxy catalogues, respectively. These estimates are remarkably close, especially for $\OmMh$,  to the true values for the MTNG models ($\sig$ = 0.816 and  $\OmMh$ = 0.142).  By fixing the value of the Hubble parameter ($h$), the constraints for our fiducial method become $\sig\ = 0.799^{+0.038}_{-0.046}$ and $0.826^{+0.040}_{-0.044}$, and $\OmMh= 0.137^{+ 0.011}_{- 0.012}$ and $0.148^{+ 0.011}_{- 0.015}$, equivalent to $\OmM= 0.298^{+ 0.024}_{- 0.027}$ and $0.323^{+ 0.025}_{- 0.032}$, respectively. The correct cosmology is thus recovered for the samples studied. Now, we will examine how different scales and different clustering statistics affect these constraints.

\begin{figure*}
\includegraphics[width=0.95\textwidth]{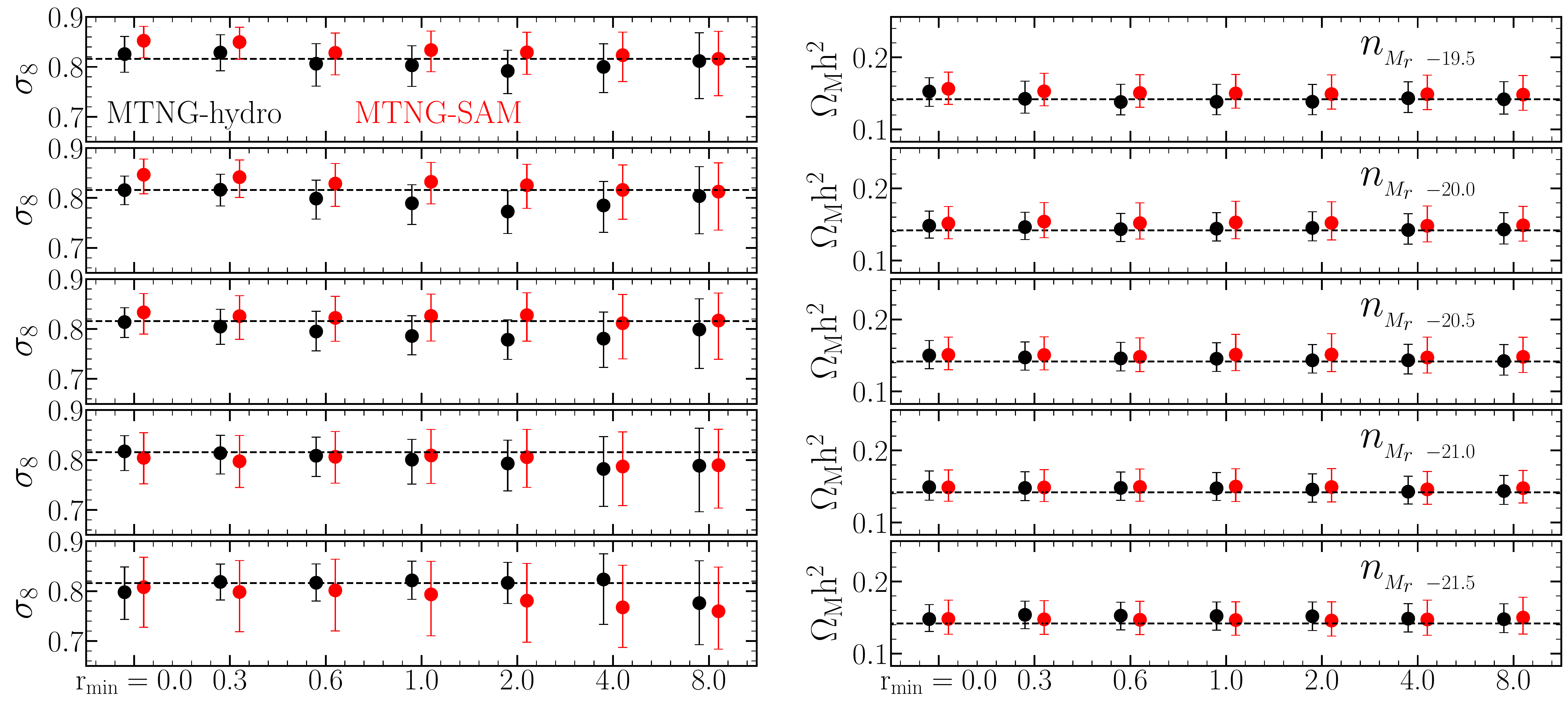}
\caption{The median (filled circles), $16^{\rm th}$ and $84^{\rm th}$ percentiles (error bars) of the $\sig$ and $\OmMh$ distributions from the MCMC chains for various $r_{\rm  min}$ values. The MTNG constraints are shown in black, while the SAM constraints are shown in red. The different rows represent different number densities, as labelled. The correct cosmology of the galaxy formation models is indicated by a dashed horizontal line as a reference.} 
\label{Fig:1D_ScaleStat_rad}
\end{figure*}

\subsection{The constraining power of different scales}
\label{sec:con_scale}

We now examine the amount of cosmological information captured by the different scales of the correlation function. This is accomplished by performing several MCMCs and limiting the minimum scale of the correlation statistics used. The scales tested are $r_{\rm  min} = 0.6$, $1$, $2$, $4$ and $8\,\hMpc$. In Fig.~\ref{Fig:1D_ScaleStat_rad} we show the median and $16^{\rm th}$ and $84^{\rm th}$ percentile distributions of $\sig$ and $\OmMh$ for the MTNG740-hydro and the MTNG-SAM galaxies. For both models, we find that for scales greater than $1-2\,\hMpc$ the constraints in $\sig$ become larger. Below this scale, the constraints are similar, except for  $\rmin=0$. At this scale, the $\chi^2$ of the best fit is quite poor, most likely because the fit or scaling in this region was insufficient to reproduce these very inner scales. For $\OmMh$, the constraints are equally good for all the values of $\rmin$, meaning that the constraining power of this parameter has low dependence on the internal galaxy distribution of each halo. It is important to notice that, independent of the value of $\rmin$, the correct cosmology is always recovered within one sigma. The only exceptions are for the highest number density sample and the two lowest $r_{\rm  min}$ values. There, we recover the correct MTNG-SAM cosmology within 1.1 $\sigma$, which are still good constraints.

\begin{figure*}
\includegraphics[width=0.95\textwidth]{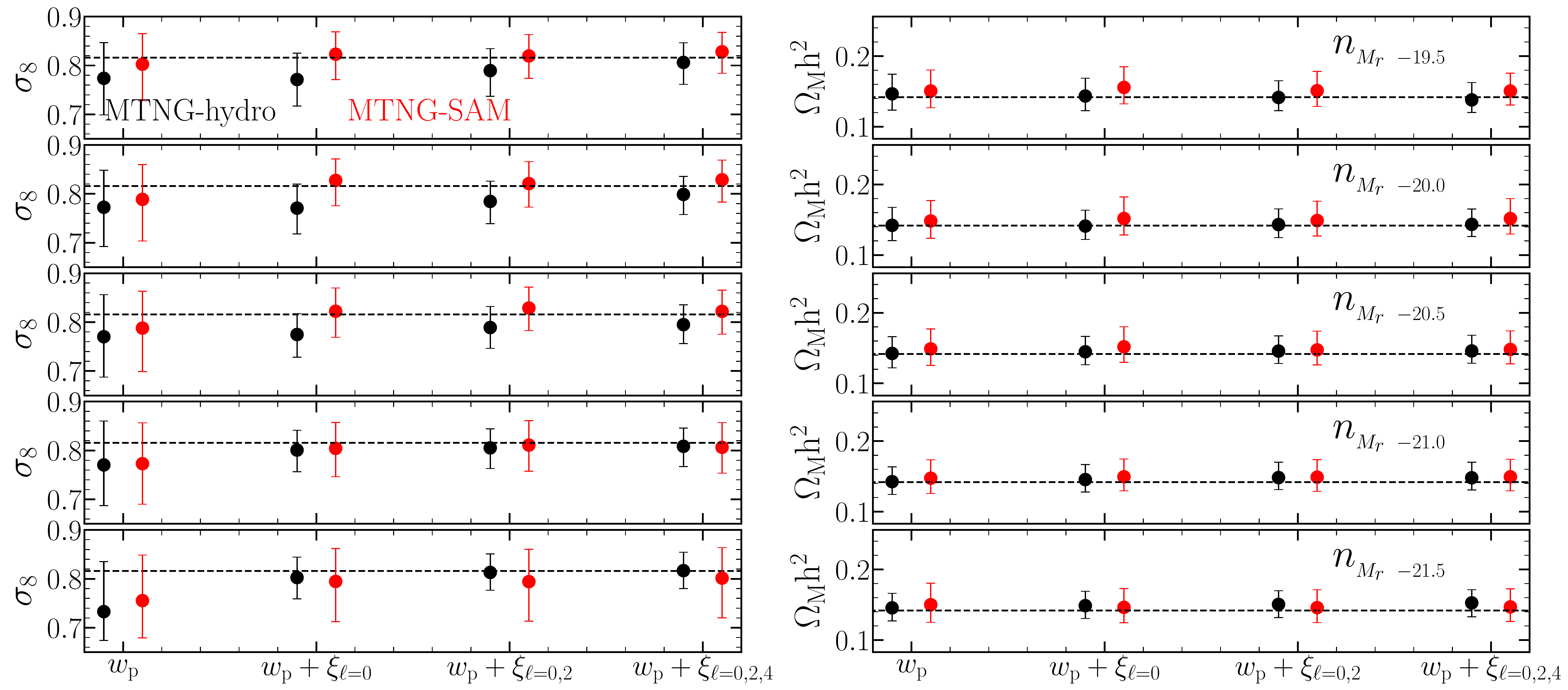}
\caption{Similar to Fig.~\ref{Fig:1D_ScaleStat_rad}, but for different statistics:\proj, \proj+\mono, \proj+\mono+\quadr, and~\proj+\mono+\quadr+\hexa} 
\label{Fig:1D_ScaleStat_stat}
\end{figure*}

\subsection{The constraining power of different clustering statistics}
\label{sec:con_stat}
We now look at the dependence of the constraints on the individual clustering statistics: the projected correlation function (\proj); the monopole of the correlation function (\mono); the quadrupole of the correlation function (\quadr); and the hexadecapole of the correlation function (\hexa). The likelihood function is computed  in the same manner as in the previous section (Eqs. 14 and 15). We use a minimal scale of $ r_{\rm  min}=0.6 \,\hMpc$, and the part of the covariance matrix that only includes \proj, \proj+\mono, \proj+\mono+\quadr,~and~\proj+\mono+\quadr+\hexa. In Fig.~\ref{Fig:1D_ScaleStat_stat} we show the median and  $16^{\rm th}$ and $84^{\rm th}$ percentile distributions of $\sig$ and $\OmMh$ for the MTNG hydro and SAM galaxies. 

For $\sig$, we find that the projected correlation function alone does not perform as well as when the monopole is included. The quadrupole and hexadecapole do not improve the constraints already achieved with the monopole. We notice that the best performance is already achieved with the projected correlation function for the lowest number density, $\OmMh$, most likely because large scales are the ones that better constrain this property, independent of the velocity profile of the galaxies, which is consistent with the previous section.

\begin{figure*}
\includegraphics[width=0.48\textwidth]{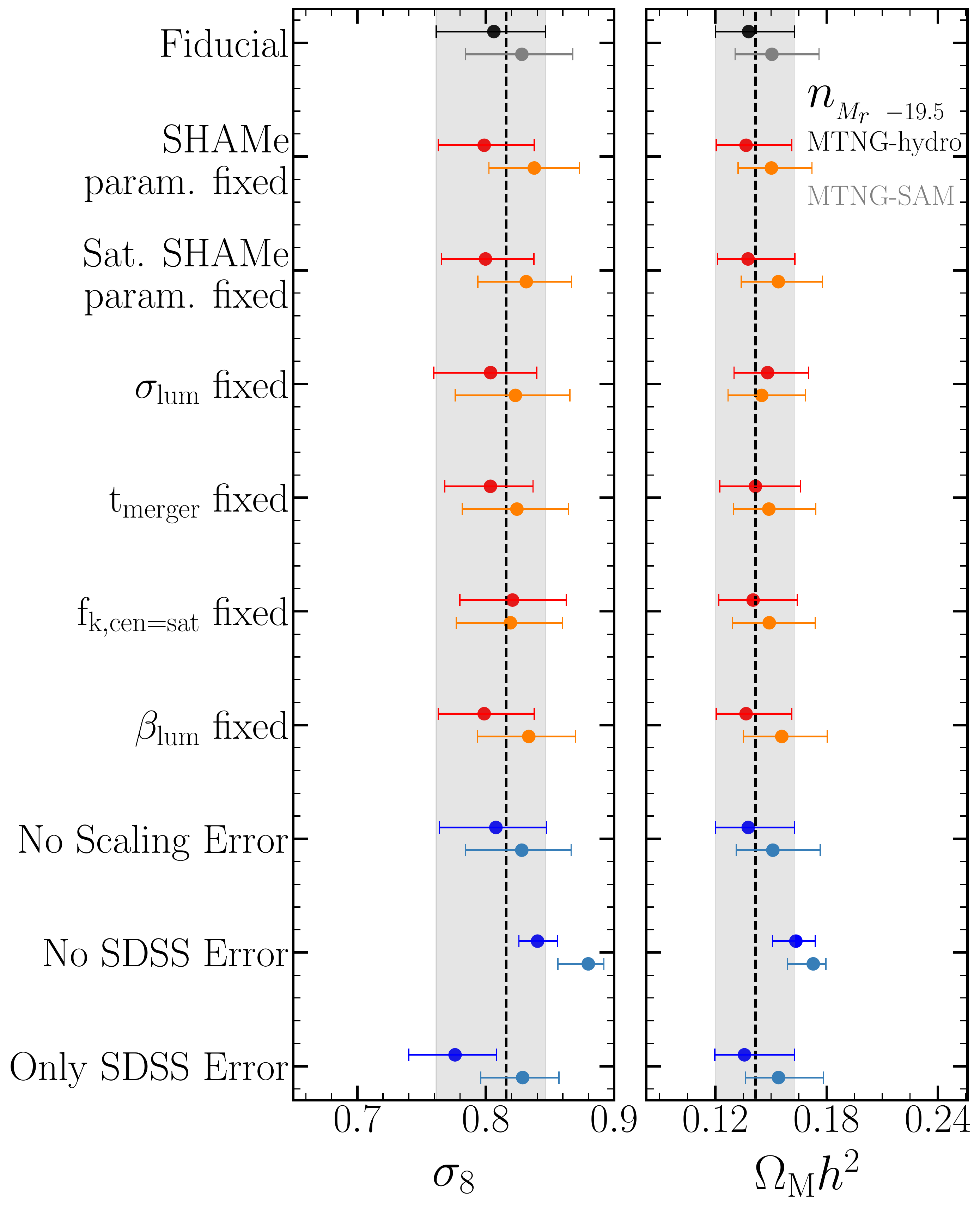}
\includegraphics[width=0.48\textwidth] {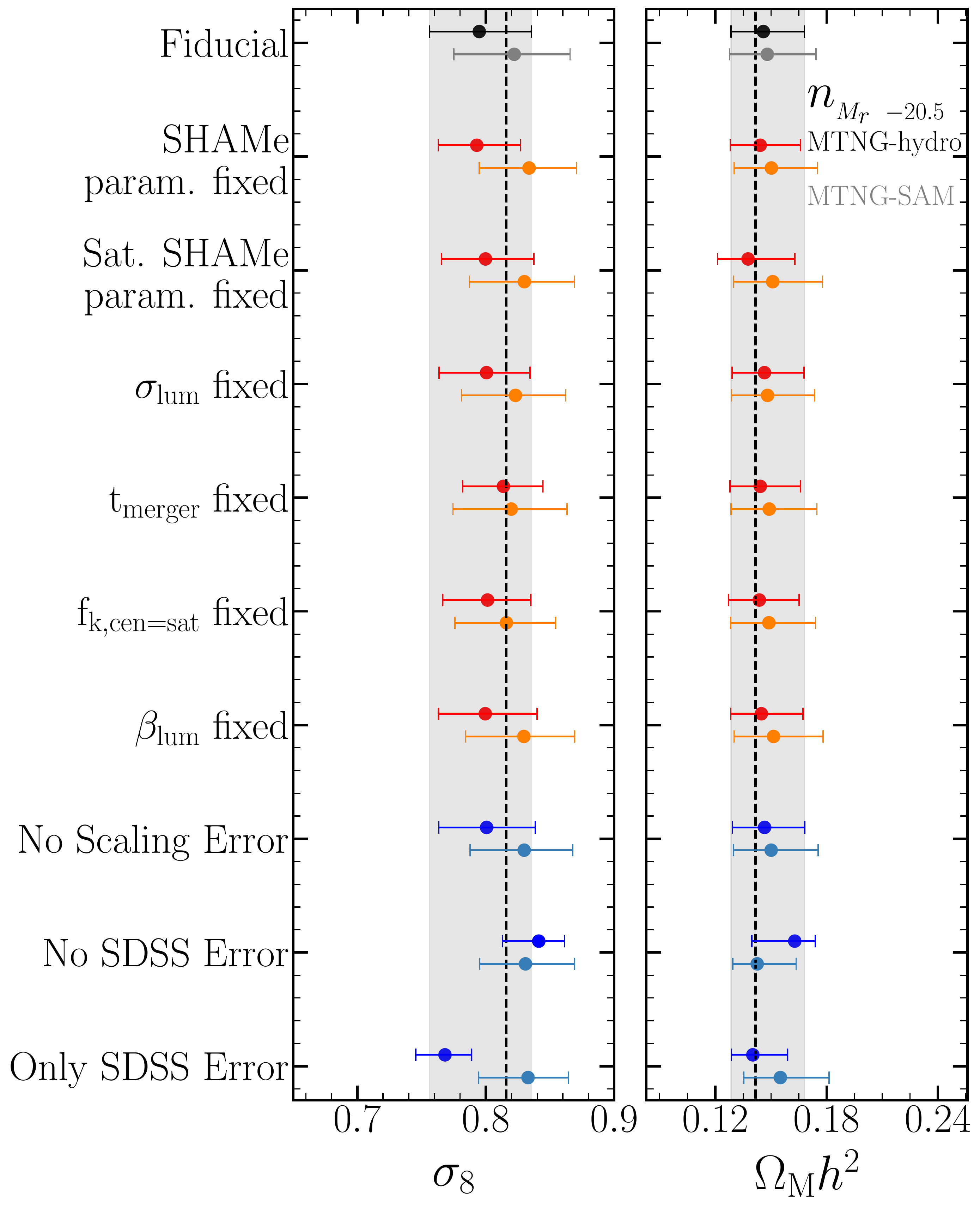}
\caption{The median (filled circles), and $16^{\rm th}$ and $84^{\rm th}$ percentiles (error bars) of the $\sig$ and $\OmMh$ distributions from the MCMC chains for various clustering statistics.  Constraints for the \Mra\ and \Mrc\ density samples are shown in the left and right panels, respectively. The darker colours represent the constraints for the MTNG-hydro galaxies, while the lighter colours represent the constraints for the MTNG-SAM. We show our fiducial case in black (using all the clustering statistics, and $ r_{\rm  min}=0.6\,\hMpc$); in red we show the constraints when fixing some or all of the different SHAMe parameters; and in blue we give the constraints when using variations of our main covariance matrix when computing the likelihood in the MCMC (see Section~\ref{sec:con_SHAMe} and Section~\ref{sec:con_cosmo} for more details). The grey shaded region denotes the percentile distribution of the MTNG's fiducial case.}  
\label{Fig:1D_post}
\end{figure*}

\subsection{Constraints on the SHAMe parameters}
\label{sec:con_SHAMe}

In the previous sections, we showed how galaxy clustering can be used to constrain cosmological information. The strength of these constraints is determined by the model's errors and the model's flexibility due to its free parameters. In this subsection, we examine how these constraints are affected by each of the SHAMe parameters, while in following subsection, we will explore how these constraints are affected by the various errors assumed for our model.

For this  subsection, we run a series of MCMC analyses with some parameters fixed to the fiducial case's best fitting parameters. The models we look at are:
\begin{itemize}
   \item {\it SHAMe parameters fixed.} We fix all SHAMe parameters, leaving free only the three cosmological parameters.
   \item {\it Sat. SHAMe parameters fixed.} We only fix the SHAMe parameters involved in the treatment of satellite galaxies, $\tmerger$ and $\betaL$.
   \item {\it $\sigL$ fixed.} We only fix the parameter that controls the scatter between the luminosity and $\vpeak$ in the SHAMe model, $\sigL$.
   \item {\it $\tmerger$ fixed.} We only fix the parameter that controls the survival rate of orphans in the SHAMe model, $\tmerger$.
   \item {\it $\Fk$ fixed.} We only fix the parameter that controls the additional level of assembly bias in the SHAMe model, $\Fk$.
   \item {\it $\betaL$~fixed.} We only fix the parameter that controls the luminosity attenuation and later disruption of the satellites in the SHAMe model, $\betaL$.
\end{itemize}

The cosmological constraints for each of these cases for the MTNG-hydro and the MTNG-SAM galaxies are shown in red colours in Fig.~\ref{Fig:1D_post}.  Constraints for the \Mra and \Mrc density samples are shown in the left and right panels, respectively. As with previous figures, the circles denote the distribution's median, and the error bars denote the $16^{\rm th}$ to $84^{\rm th}$ percentiles. The shadow region corresponds to the $16^{\rm th}$ to $84^{\rm th}$ percentiles of the MTNG's fiducial case.

We find no significant improvement in the constraints of $\sig$ and $\OmMh$ by fixing any of the SHAMe parameters. This is consistent with the finding that there is no significant correlation between the cosmological parameters and the SHAMe parameters (Fig.~\ref{Fig:full_mcmc}). These results indicate that none of the SHAMe free parameters reduced the constraining power on the cosmological parameters. This does not imply, however, that a more realistic model would not aid in enhancing the constraints. As will be shown in the following section, the different components of the covariance matrix, including the component that accounts for the error of the SHAMe model, will have an impact on the overall constraint of our model.

\subsection{The impact of the errors on the cosmological constraints}
\label{sec:con_cosmo}

Finally, we measure how the constraints depend on the systematics of our model. In Section~\ref{sec:errsum} we showed that the largest sources of error in the combined covariance matrix are cosmic variance (assumed to be identical to that in the SDSS) and the scaling technique. We measured the impact of these systematics by running three MCMC analyses with different covariance matrices:
\begin{itemize}
   \item {\it No Scaling Error.} Same as our combined covariance matrix, but without the scaling component of the error.
   \item {\it No SDSS Error.} Same as our combined covariance matrix, but without the cosmic variance component of the error.
   \item {\it Only SDSS Error.} Only use the SDSS covariance matrix.
\end{itemize}
The constraints for these three runs and for the MTNG-hydro and the MTNG-SAM are depicted in blue colours in Fig.~\ref{Fig:1D_post}. While scaling is the second largest source of error, ignoring its error has a negligible effect on the cosmological constraints. We would like to remind the reader that the scaling technique is capable of quickly ``producing'' a dark matter simulation for a given cosmology in a few seconds. This technique is the cornerstone of our approach, and demonstrating that its associated error has a negligible effect on the cosmological constraints is important to validate our approach.

Now we focus on the assumed cosmic variance error, which was taken from the SDSS. This error was included for comparison purposes only, and is not part of our model's intrinsic error. By removing this error from our covariance matrix, we find stronger constraints for $\sig$, but not for $\OmMh$. These fiducial constraints represent the full potential of our approach in its current state. If, hypothetically speaking, we had access to a galaxy survey with perfect clustering measurements and ran our model over much larger simulations, the cosmological constraints would look similar to these. It is worth noting that the median estimate of $\sig$, which is a bit lower for the fiducial case than the correct value, is now much larger for both number densities. This shift to a lower value in the fiducial case may be due to the SDSS covariance matrix's larger errors at large scales (Fig.~\ref{Fig:full_err}), or it may be induced by the specific cross-correlation between different scales and statistics. This shift in the value of $\sig$ should be kept in mind when interpreting constraints derived from this covariance matrix.

Finally, we look at the constraints when only cosmic variance errors are included. Here, we do not detect significant changes in the size of the error bars for either $\sig$ or $\OmMh$, particularly at lower number densities. Nonetheless, we notice that the constraints on $\sig$ do not completely enclose the MTNG's correct value. Interestingly, the value predicted is lower than the real value. This highlights the importance of including model errors (e.g. in the galaxy population model, in the emulator, etc.) in these kinds of studies, specially considering that some observational studies (eg. \citealt{Nunes:2021,Yuan:2022}) are finding low values for $S_8 = \sig \sqrt{\OmM/0.3}$ and f$\sig$, with ``f'' representing the growth rate of structure obtained from linear perturbation theory. 

We conclude from these two last sections that the constraints found in our fiducial model could be improved further by using more powerful observational data for $\sig$, and a more realistic galaxy population model for $\OmMh$.

\begin{figure*}
\includegraphics[width=0.95\textwidth]{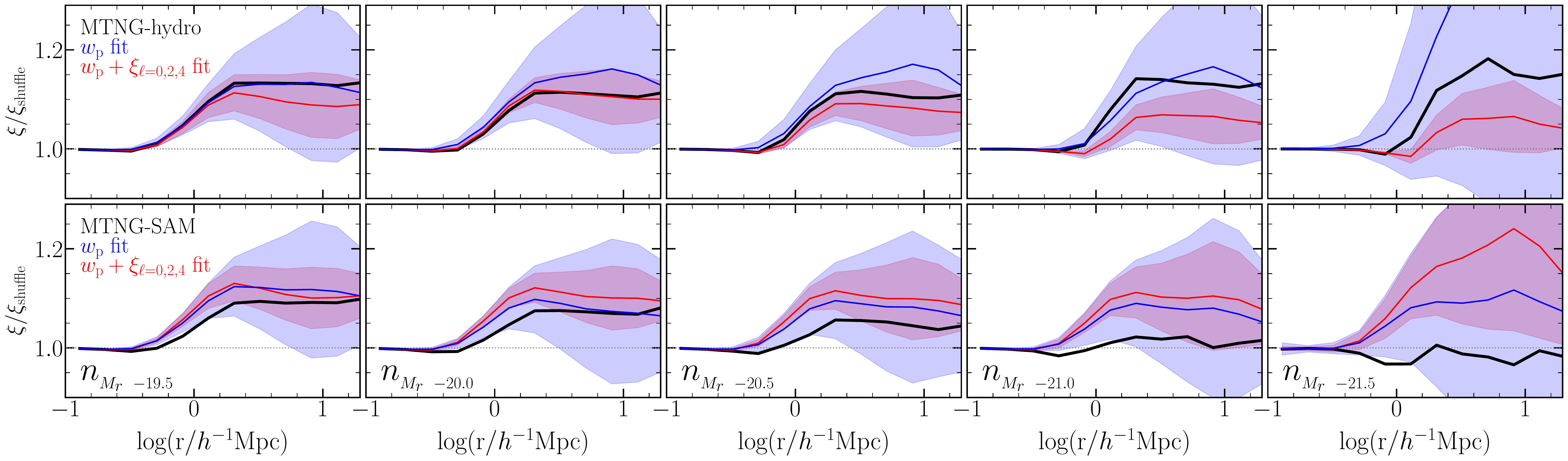}
\caption{
The predicted galaxy assembly bias signal of the SHAMe mocks in comparison to the actual assembly bias signal of the MTNG-hydro (top panel) and the MTNG-SAM (bottom panel) galaxies. In blue and red we show the estimated assembly bias signal when fitting only the projected correlation function, and when fitting all the clustering statistics of the galaxy formation models, respectively. The different columns represent different number density samples, as labelled. For the galaxy formation models, the galaxy assembly bias signal is computed by averaging the ratio of their correlation functions across 20 different shuffling runs (see Section~\ref{sec:AB} for more details). For the mocks, we took 200 random points from the MCMC chains, recreated the equivalent mocks, and computed the assembly bias similarly as we did for the galaxy formation models (although, with only 5 shufflings per mock). The solid lines and shaded regions represent the mean and the $16^{\rm th}$ and $84^{\rm th}$ percentiles of each distribution, respectively.}  
\label{Fig:GAB} 
\end{figure*}

\section{The constraints on assembly bias}
\label{sec:AB}

As mentioned before, our SHAMe model has one parameter, $\fk$, which is responsible for regulating galaxy assembly bias. When omitting this parameter, the value of $\rm \chi^2/d.o.f.$ for the best-fitting model increases (not shown here). This means that including this parameter contributes to a more accurate reproduction of galaxy clustering. This is why the parameters associated with galaxy assembly bias (occasionally referred to as bias parameters) are included in the fitting of other galaxy occupation models such as HODs. While several studies have included an assembly bias parameter in galaxy clustering models in order to constrain cosmological parameters  (e.g. \citealt{Yuan:2022}) or galaxy assembly bias (e.g. \citealt{Salcedo:2021}), to our knowledge, no one has examined directly the effect of assembly bias uncertainties when estimating cosmological parameters. 

We are interested in determining whether the level of assembly bias estimated by our methods matches the true level in the underlying galaxy formation models for two reasons: (1) To know whether $\fk$ is helping to fit galaxy clustering by correctly adding assembly bias (as intended), or, for example, by unrealistically changing assembly bias to compensate for some limitation of the basic SHAMe approach; (2) To determine whether we can use our methods to constrain not only cosmological parameters but also assembly bias itself.

To quantify how well our model constrains assembly bias, we randomly select 200 points from the MCMC chains (after the burn-in) and compute the assembly bias of these samples, using the shuffling technique introduced by \cite{Croton:2007}. In this technique, a ``shuffled mock'' is created by shuffling the galaxy population between haloes of similar mass. This is done in bins of 0.1 dex in $\log(h^{-1}M_{\odot})$. In previous works, we have checked alternative values for this binning, finding similar results. By construction, since the galaxy occupation of this shuffled mock does not depend on the halo mass, it has no galaxy assembly bias. Galaxy assembly bias is then defined as the square of the ratio between the two-point correlation functions of the original and the shuffled runs:
\begin{equation}
\xi/\xi_{\rm shuffle} = b^2.
\end{equation}

For each of the 200 points of the MCMC we compute 4 distinct mocks, each with a unique random seed in each pair of scaled simulations (we need to rescale the simulations in the same way we did to create the training points of the emulator). For each mock, we produce 5 shuffled mocks and compute their correlation function. We repeat this for each number density, both for the MTNG-hydro and MTNG-SAM MCMC chains, and for runs done only with the projected correlation function (usually used to constrain galaxy assembly bias) and also including the multipoles of the correlation function. This is equivalent to creating 800 standard mocks and 4,000 shuffled mocks. 

For each sample, we compute the mean and standard deviation of the 200 correlation function ratios, which we characterize as the constrained assembly bias from our mocks. We verify that, even with 50 points, we have a reasonably robust measure of the assembly bias signal's mean and standard deviation. We compare these ratios with those of the actual MTNG-hydro and MTNG-SAM galaxies in Fig.~\ref{Fig:GAB}. The assembly bias of the galaxy formation simulations was computed in a similar way as for the SHAMe catalogues. Due to the fact that we only have one realization of the galaxy formation simulations, we used twenty shuffled runs to further reduce measurement noise. We would like to remind the reader that the MCMCs from these points were used to fit the various galaxy clustering statistics but not this galaxy assembly bias signal.

When doing the fitting only using the projected correlation function, we can recover the correct galaxy assembly bias signal within one sigma at all number densities. When including the multipoles of the correlation function, we can only recover the assembly bias signal within one sigma for the highest number densities. For the two lowest number densities, we find some systematic differences of up to 1.5 sigma.

The differences found when we include the multipoles in the least number density sample could be due to a number of factors, including: (a) Some limitations of the SHAMe model (e.g.~the lack of velocity bias), which may not be very necessary for cosmological constraints but needed for assembly bias constraints; (b) That the shuffling of the MTNG740 galaxies and the SAM galaxies, which has a larger resolution than the simulation the SHAMe models and the SAM were run on, along with the halo finder algorithm used, generates a different shuffling run (lead by the different classification of splashback galaxies, which can be responsible for part of the galaxy assembly bias signal, Zehavi et al. in prep.), predicting a different galaxy assembly bias signal. For the {\small L-GALAXIES} semi-analytical model, which was run on a dark matter simulation with the same resolution as the one used by the SHAMe model, we can recover the right level of assembly bias for all number densities (not shown here).

Even in this latter case, we confirm that the assembly bias levels estimated by our procedures are consistent with those present in the original galaxy formation models, at least for the higher number density samples. 
Additionally, we demonstrate that SHAMe model combined with the scaling technique has the potential to estimate galaxy assembly bias on high number density samples. Further, we would like to encourage other groups to infer assembly bias from the reconstructed models of their MCMC chains, rather than constraining their model's galaxy assembly bias parameters, which are not easily interpretable in terms of the actual assembly bias signal.

\section{Summary}
\label{sec:Summary}
In this work, we have developed a new method for estimating cosmological parameters from the redshift-space clustering of galaxies. The N-body scaling technique can rapidly generate a simulation assuming some requested cosmology based on an available simulation assuming a nearby but different cosmology;  the SubHalo Abundance Matching extended model (SHAMe) is a physically motivated empirical model capable of reproducing galaxy clustering in real- and redshift-space. Combining them, we generate over 175,000 clustering measurements for various cosmological and SHAMe parameters. This allows us to construct an emulator capable of reproducing galaxy clustering (\proj,~\mono,~\quadr~and~\hexa)  in a fraction of a second. 

With this emulator, an MCMC analysis is able to estimate cosmological parameters from galaxy clustering data while marginalising over galaxy formation uncertainties. We test this procedure using the MTNG740 hydrodynamic simulation, and various {\small L-GALAXIES} semi-analytic models applied to one of the MTNG740-DM simulations. We test how resulting constraints on cosmological parameters depend on the scales used in the clustering analysis, on the clustering statistics included, and on the kind of errors accounted for. Here we repeat our most important findings:

\begin{itemize}

\item The projected correlation function, monopole, quadrupole and hexadecapole are more sensitive to changes in $\OmM$, $\sig$ and $\h$ than the rest of the cosmological parameters (Fig.~\ref{Fig:derivative}). The next most relevant parameters are the neutrino mass $\Mnu$ and $\ns$. In the future, we plan to study how cosmology impact other statistics, such as 3PCF \citep{Guo:2016b}, kNN-CDF \citep{Banerjee:2020}, lensing, etc. This could potentially help us improve the constraints of the remaining cosmological parameters.

\item We measured the errors coming from the SHAMe model, the scaling and the emulator, finding that the error of the scaling dominates (Fig.~\ref{Fig:SHAMe_err}, Fig.~\ref{Fig:Emu_Err} and Fig.~\ref{Fig:Scale_err}). These errors are still, in most cases, lower than the statistical sampling error we would get from the SDSS (Fig.~\ref{Fig:full_err}). Finally, we find that, while the error from the scaling technique is considerable, it does not have a strong impact on constraints on the cosmological parameters (Fig.~\ref{Fig:1D_post}).

\item By running an MCMC with our emulator using a covariance matrix which combines all our error estimates, we obtain $\sig\ = 0.799^{+0.039}_{-0.044}$ and $0.826^{+0.041}_{-0.045}$, and $\OmMh= 0.138^{+ 0.025}_{- 0.018}$ and $0.151^{+ 0.026}_{- 0.020}$ from the clustering of a \Mra density sample of the MTNG-hydro and the MTNG-SAM galaxies, respectively. In each case, these are very close to the true values ($\sigma_{8,{\rm MTNG}} =$ 0.8159 and $\Omega_{\mathrm{M}} h^2_{\rm MTNG} =$  0.142). 

\item The cosmological constraints for $\sig$ improve significantly when we start using scales below $2\, \hMpc$ for the MTNG-hydro galaxies. For the current MTNG-SAM, and for $\OmMh$, we do not find a major improvement when using smaller scales (Fig.~\ref{Fig:1D_ScaleStat_rad}).

\item The projected correlation function is good enough to constrain $\OmMh$, but the monopole is needed to obtain good constraints on $\sig$. The quadrupole and hexadecapole do not significantly improve what is already achieved using projected correlations and the monopole (Fig.~\ref{Fig:1D_ScaleStat_stat}).

\item When running an MCMC with the SHAMe parameters fixed, the constraints on $\OmMh$ and $\sig$ do not vary significantly (Fig.~\ref{Fig:1D_post}). This suggest that none of the SHAMe free parameters reduced the constraining power on the cosmological parameters.

\item The constraints on $\sig$ tighten when running an MCMC with a covariance matrix excluding the error coming from cosmic variance (taken at present from the SDSS). This means that our current method would yield improved constraints if we applied it to larger and more powerful galaxy surveys.

\item While the error from cosmic variance is the largest source of uncertainty, failing to include the rest of the errors in our procedures causes bias in our constraints. This highlights the importance of accounting for all known modeling errors in studies of this kind.

\item We have demonstrated that the level of assembly bias estimated by our procedures is consistent with that of the underlying galaxy formation model, suggesting that our methodology can be used to estimate galaxy assembly bias from observational samples (Fig.~\ref{Fig:GAB}).

\end{itemize}

We would like to emphasise that, while the constraints identified in this work are tight, its primary accomplishment is to demonstrate that these constraints are indeed realistic. In future work, we will use this methodology to constrain cosmological and galaxy formation parameters based on SDSS clustering statistics. Beyond this, in preparation for the next generation of galaxy surveys, we will expand the number of statistics we employ in order to constrain additional cosmological parameters and to move to other types of galaxy sample, for example, emission line galaxies.

\section*{Acknowledgements}

We thank Hong Guo for some useful clarifications, Daniele Spinoso for assistance when running {\small L-GALAXIES} and Jon\'as Chaves Montero for some useful discussions.  SC acknowledges the support of the ``Juan de la Cierva Incorporac\'ion'' fellowship (IJC2020-045705-I).
REA acknowledges the support of the ERC-StG number
716151 (BACCO) and the Project of excellence Prometeo/2020/085 from the Conselleria d'Innovaci\'o, Universitats, Ci\`encia i Societat Digital de la Generalitat Valenciana. CH-A acknowledges support from the Excellence Cluster ORIGINS which is funded by the Deutsche Forschungsgemeinschaft (DFG, German Research Foundation) under Germany's Excellence Strategy -- EXC-2094 -- 390783311. VS and LH acknowledge support by the Simons Collaboration on ``Learning the Universe''. LH is supported by NSF grant AST-1815978.  SB is supported by the UK Research and Innovation (UKRI) Future Leaders Fellowship [grant number MR/V023381/1].  The authors also acknowledge the computer resources at MareNostrum and the technical support provided by Barcelona Supercomputing Center (RES-AECT-2019-2-0012 \& RES-AECT-2020-3-0014). The authors gratefully acknowledge the Gauss Centre for Supercomputing (GCS) for providing computing time on the GCS Supercomputer SuperMUC-NG at the Leibniz Supercomputing Centre (LRZ) in Garching, Germany, under project pn34mo. Finally, we would like to thank our referee, Johannes Ulf Lange, for his careful reading and helpful comments throughout the review process.

\section*{Data Availability}
The data underlying this article will be shared upon reasonable request to the corresponding author. The MillenniumTNG simulations will be made fully publicly available at \url{https://www.mtng-project.org/} in 2024. 

\bibliographystyle{mnras}
\bibliography{Biblio}

%\appendix

\bsp
\label{lastpage}
\end{document}